# TWO-GAP SUPERCONDUCTOR ZrB$_{12}$ WITH DYNAMIC STRIPES AND CHARGE DENSITY WAVES: CRYSTAL STRUCTURE, PHYSICAL PROPERTIES AND PAIRING MECHANISM

A. N. Azarevich[a], N. B. Bolotina[b], O. N. Khrykina[b], A. V. Bogach[a], K. M. Krasikov[a],
A. Yu. Tsvetkov[c], S. Yu. Gavrilkin[c], V. V. Voronov[a], S. Gabani[d], K. Flachbart[d],
A. V. Kuznetsov[e], N. E. Sluchanko[a,*]

[a]*Prokhorov General Physics Institute of RAS, 119991 Moscow, Vavilov str.,38, Russia*
[b]*National Research Center Kurchatov Institute, 123182 Moscow, Academician Kurchatov sq., 1, Russia*
[c]*Lebedev Physical Institute of RAS, 119991 Moscow, Leninskiy Av. 53, Russia*
[d]*Institute of Experimental Physics of the SAS, SK−04001 Košice, Watsonova, 47,Slovakia*
[e]*National Research Nuclear University (MEPhI), 115409, Moscow, Kashirskoe sh., 3, Russia*

*e-mail: nes@lt.gpi.ru

A review of long-term studies of ZrB$_{12}$ and LuB$_{12}$ superconductors with very similar conduction bands and phonon spectra, but with radically different (by a factor of 15–20) critical temperatures and magnetic fields is presented. A detailed analysis of well-known studies in combination with new results of structural, thermodynamic and charge transport measurements obtained here for these metallic dodecaborides with Jahn-Teller instability of the rigid boron network and with dynamic charge stripes allows us to conclude in favor of the primary role of nanoscale effects of electron phase separation, leading to the formation of one-dimensional dynamic chains with different configurations of fluctuating charges, which in the case of ZrB$_{12}$ are predominantly 2p states, and for LuB$_{12}$ - 5d-2p states. We propose a new *plasmon-phonon pairing mechanism* in ZrB$_{12}$, which may be common to different classes of high-$T_c$ superconductors.

## INTRODUCTION

Transition metal (TM) dodecaboride ZrB$_{12}$ is a very hard metallic compound that goes into a superconducting state at a critical temperature $T_c \approx$ 6 K, which is more than an order of magnitude higher than $T_c$ of about 0.4 K for relative rare-earth (RE) dodecaboride LuB$_{12}$ [1-3]. Like most TM and RE dodecaborides, ZrB$_{12}$ crystallizes in the face centered cubic (*fcc*) lattice (Fig. 1) and corresponds well to the $Fm\bar{3}m$ symmetry group. At room temperature 293 K, the lattice parameter of ZrB$_{12}$ is $a$ = 7.4075(1) Å [4]. The structure is formed by negatively charged [B$_{12}$]$^{2-}$ cuboctahedrons and Zr$^{4+}$ cations, alternating like Na and Cl in the NaCl structure. Each metal ion is surrounded by 24 boron atoms, which form an octahedron with truncated vertices. Two 5$s$ electrons of tetravalent Zr are incorporated into the B$_{12}$ electron configuration to provide the valence balance [5, 6], while two 4$d$ electrons remain in the conduction band and are responsible for charge transport. The calculations of the band structure of ZrB$_{12}$, first reported by Shein & Ivanovskii [7] and then confirmed by Jäger et al. [8], indicate a fairly wide conduction band (~ 2 eV) in the cage-cluster compound. The electronic structure and bulk properties of ZrB$_{12}$ were studied by Grechnev et al. [9] using *ab initio* calculations supplemented by magnetic and ultrasonic measurements.

As shown by Lortz et al. [10] the phonon spectra of ZrB$_{12}$ contains both very high energy (up to 140 meV, see also [11]) and low energy (~15-20 meV) branches. A comparative analysis of the heat capacity, resistivity, and thermal expansion results allows concluding that low-energy vibrations (~17.5 meV) of Zr-ion [11, 12], which play an important role in the superconductivity of ZrB$_{12}$, are anharmonic and associated with changes in the electronic system of crystal. A similar situation was described by Mandrus et al. [13] for LaB$_6$. There was no consensus in the beginning as to whether the vibrations of the cations could actually be described by the Einstein model either as a single-phonon mode or as several separate phonon modes [10], or there was a phonon distribution that could be modeled using the Debye spectrum [14]. The results of Glushkov et al. [15], who studied the Seebeck effect on ZrB$_{12}$ single crystals, supported the scenario with the Einstein oscillators. These rattling quasi-local modes corresponding to quasi-independent vibrations of cations were also detected in various dodecaborides in the energy range of 14–17.5 meV in measurements of inelastic neutron scattering [11, 12, 16, 17] and heat capacity [18, 19].

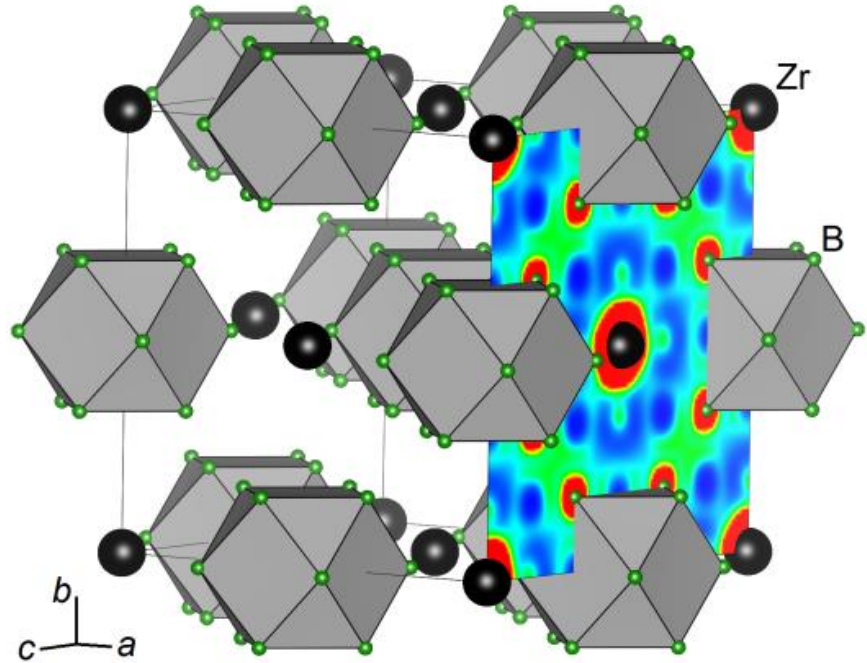

Fig. 1. Schematic representation of the NaCl-type unit cell of $ZrB_{12}$. The dark balls are Zr atoms, the small green balls are B atoms, forming cuboctahedrons $B_{12}$. The map of the electron density distribution in the cell face was obtained using the maximum entropy method (see text).

To understand why $ZrB_{12}$ becomes a superconductor at a $T_c$ about 15 times higher than $LuB_{12}$, Teyssier et al. [20] studied in detail specific heat, electrical resistivity, and optical spectroscopy in these two compounds. Extracting superconducting properties, phonon density of states, and electron-phonon coupling (Eliashberg function) from the above measurements these authors discussed the factors that could govern $T_c$ trying to explain the difference between the two compounds. They concluded that the main reason for the lower $T_c$ in $LuB_{12}$ was the weaker coupling of the Einstein phonons to the conduction electrons. The origin of this weaker electron-phonon interaction remained unclear but might be found in the different "volume filling factors" of the boron cages. This factor, which is defined by Teyssier et al. [20] as the ratio between the volumes of the metal ion and the $B_{24}$ polyhedron, tunes the hybridization of the $d$- orbitals of Lu/Zr ions with those ($2p$) of the boron framework and therefore strongly influences the electron-phonon interaction.

Previously, Werheit et al. [21] studied the Raman spectra of $ZrB_{12}$ and $LuB_{12}$ single crystals and came to the conclusion that a small fraction of metal atoms were slightly displaced from the centers of the $B_{24}$ polyhedrons. This explained the occurrence of rather strong Raman lines in the low-frequency spectral range, where vibrations of heavy atoms of metal were detected. The idea of partial disordering of the cation lattice in dodecaborides was caught up by Teyssier et al. [22], who analyzed the optical properties of a $ZrB_{12}$ single crystal in the temperature range of 20 - 300 K, measured by a combination of optical reflectivity and ellipsometry. The optical experiments showed an anomalous temperature dependence of the plasma frequency in $ZrB_{12}$. With reference to Werheit et al. [21], the above authors calculated the band structure and plasma frequency as a function of the possible displacements of the Zr-ion in the [100], [110] and [111] directions. Assuming the reality of the displacements, they could explain the anomaly in the temperature dependence of the plasma frequency. Later, a similar anomaly was discovered by Teissier et al. [20] in $LuB_{12}$, which could lead to a stronger suppression of the density of electronic states at the Fermi level at low temperatures.

The next step in the interpretation of the lattice disordering in dodecaborides was taken by Sluchanko et al. [23-26], who measured the heat capacity and Raman spectra in $Lu^NB_{12}$ and $Zr^NB_{12}$ single crystals with different boron isotopes (N = 10, 11, nat) at low and intermediate temperatures. Hereinafter, "nat" means natural isotope composition, with a ratio of $^{11}B:^{10}B \approx 4:1$. The low-temperature anomalies in the specific heat, along with the features of the Raman spectra, were interpreted as follows. The main and displaced positions of a cation are separated by a barrier in a double-well potential (DWP) of height $\Delta E$. At temperatures above $T^* = \Delta E / k_B$, the cationic lattice is dynamically disordered, but below this temperature the cations "freeze" in one of the DWP minima, unable to overcome the potential barrier, which leads to static disorder of Zr/Lu ions. The crystal transforms into a "cage-glass" state, which is a mixture of crystal (rigid covalent boron framework) and glass (disordered clusters of displaced heavy TM/RE ions). This transition leads to renormalization of the low-frequency vibration spectra and the appearance of a boson peak in the Raman spectra, as well as to Schottky anomalies in the low-temperature heat capacity associated with the emergence of two-level systems (TLS) in the disordered structure of $LuB_{12}$ and $ZrB_{12}$ in the presence of boron vacancies. It has been suggested that the same transition may be associated with the unusual temperature behavior of the Hall coefficient and magnetic susceptibility of $Lu^NB_{12}$ (N = 10, 11, nat) near the corresponding temperature $T^*$ in the temperature range 50–70 K [27]. It can also be the reason for the previously observed anomalies of charge transport [28] and magnetoresistance [29] in several $RB_{12}$ dodecaborides ($R$ = Ho, Er, Tm, Lu).

Like other highly symmetric molecules, such as $B_{12}$ icosahedrons in boron and metallic higher borides [30, 31], $B_{12}$ cuboctahedrons in dodecaborides undergo Jahn – Teller (JT) distortions [32]. Due to the interactions between the nearest $B_{12}$ clusters in the $RB_{12}$ crystal, local distortions become mutually consistent, which leads to a small static, and a significant dynamic distortion of the lattice geometry. This phenomenon is known as the cooperative JT effect and is well described in the literature [33-36]. Cooperative interactions in $RB_{12}$ can lead either to a parallel alignment of all local distortions of the $B_{12}$ cuboctahedra (so-called ferrodistortive case) or to a more complex geometric arrangement of local $B_{12}$ distortions (antiferrodistortive effect). The most likely scenario for both TM and RE dodecaborides is the formation of a ferrodistortive phase with long-range ordering of JT distortions and a very small static deformation of the *fcc* lattice, which was reported in [37-42], for recent review see [43]. The *static* component of the JT effect does not require a revision of the structural model, but the *dynamic* JT instability manifests itself in the marked anisotropy of the charge transport parameters in directions <110> that should be equivalent in a cubic crystal (for $LuB_{12}$ see, for example, [44]). Indeed, it was shown in [40] that the delocalized fluctuating part of the electron density in $LuB_{12}$, which cannot be taken into account in the structural model, is unevenly distributed along symmetry-equivalent <110> directions in the crystal, which correlates very accurately with the anisotropy of the charge transport. It was discussed in [43] that the *dynamic* component of the JT effect in $RB_{12}$ is responsible for periodic changes of the *4d(5d)-2p* hybridization of TM (RE) and boron electron states leading to modulation of the conduction band width and formation of the *dynamic charge stripes*. Additionally, formation of vibrationally coupled pairs of RE ions was deduced from the ESR, heat capacity and structural studies of $YbB_{12}$ [45, 46], and it was observed in $HoB_{12}$ [47] in the heat capacity, magnetization and inelastic neutron scattering measurements. Note, that both static and dynamic JT components are very important for electron-phonon interaction, Cooper pairing and the pair breaking in superconductors.

Recently, detailed studies of superconducting and normal state characteristics were carried out at low temperatures in $Lu_xZr_{1-x}B_{12}$ solid solutions [48]. It was proposed in [48] that charge fluctuations at the Lu sites cause a pair-breaking effect, which leads to a 15-fold decrease in $T_c$ with increasing $x$. As shown in [49, 48, 50], $ZrB_{12}$ is a two-gap strongly coupled *s-wave* superconductor in the dirty limit, where the charge stripes are not linear as detected for $LuB_{12}$, and at least at $T \sim 100$ K these form *two grids from rhomboid cells (checkerboard patterns)* built from (*i*) hybridized 4d-2p and (*ii*) only 2p conduction band states of $ZrB_{12}$. In [51] authors also argued in favor of two-band type-1.5 superconductivity in $ZrB_{12}$. On the contrary, in recent study [52] an anisotropic superconductivity near the critical Bogomolnyi point (Ginzburg-Landau-Maki (GLM) parameter $\kappa_c \approx 0.7$) with field-direction-dependent type-I and type-II behavior was discussed. Thus, currently the possible reasons for the $T_c$ increase and variation of the electron-phonon interaction $\lambda_{e\text{-}ph}$ in the range 0.4–1 in $Lu_xZr_{1-x}B_{12}$ are not yet clear. The two-gap superconductivity in $ZrB_{12}$ was confirmed also in [50, 53] in the detailed heat capacity studies, which detected two superconducting phase transitions at least for magnetic field directions $H \| [110]$ and $H \| [111]$. Additionally the sub-structural charge density wave (s-CDW) was detected at low temperatures [53] and pseudo-gap was established in the XPS experiments in $ZrB_{12}$ [54]. Taking into account that such ingredients as (1) dynamic charge stripes, (2) pseudo-gap and (3) CDW all observed above $T_c$, and (4) multiband superconductivity are considered usually as the fingerprints of the unconventional high-$T_c$ superconductivity (HTSC) in cuprates and Fe-based pnictides, where the pairing of electrons is mediated likely by spin-fluctuations [55], it seems promising to study in more detail the features of electron-phonon pairing in the "conventional" superconductor $ZrB_{12}$, where singularities (1)-(4) have been reliably detected.

In addition, we also note the similarity of the crystal structure (*fcc* lattice, space group $Fm\bar{3}m$) of $ZrB_{12}$ and cage-cluster polyhydrides $(La,Y)H_n$ (n=10), where HTSC with the highest values of $T_c \geq 250$ K has currently been discovered at pressures up to 170 GPa, and the electron-phonon pairing mechanism is widely discussed (see, for example, [56-58]). Assuming the development of dynamic Jahn-Teller structural instability of $H_n$ clusters in $(La,Y)H_n$ under quasi-hydrostatic pressures, one can expect a number of features of the implementation of the electron-phonon mechanism of superconductivity, common to "unconventional" HTSC polyhydrides $(La,Y)H_n$ and "conventional" superconductor $ZrB_{12}$.

## EXPERIMENTAL DETAILS

The details of preparation of the $ZrB_{12}$ single crystals are given in [21, 48, 59]. Samples #1 – #4 of $ZrB_{12}$, intended for measurements of magnetoresistance (MR) and Hall effect (HE), were cut in the form of rectangular plates of about $0.3 \times 0.4 \times 4$ mm$^3$ in size from the same single crystal rod. The long sides of the plates were oriented along the crystallographic directions <110> or <100> with accuracy ~2°. The angular dependences of the resistivity and Hall voltage were measured using an original method of sample rotation with step-by-step fixation of the sample position in a steady magnetic field, step $\Delta\varphi=1.8°$, angle $\varphi \equiv \mathbf{n}^{\wedge}\mathbf{H}$, where $\mathbf{n}$ is the normal vector to the lateral surface of the rectangular sample, see the inset to Fig. 12 below. The magnetic field $\mathbf{H}$ with strength up to 80 kOe supplied by superconducting magnet was applied perpendicular to the measured direct current (DC) $\mathbf{J} \parallel$ <100> in #1-#3 and $\mathbf{J} \parallel$ <110> in #4 (Fig. 12 below). The installation equipped with a step-motor with automated control of the step-by-step sample rotation is similar to that applied earlier in [60]. High accuracy of the temperature control ($\Delta T \approx 0.001$ K in the range 1.8 – 7 K) and magnetic field stabilization ($\Delta H \approx 2$ Oe) was achieved by using the commercial temperature controller TC 1.5/300 and superconducting magnet power supply unit SMPS 100 (Cryotel Ltd.) in combination with CERNOX 1050 thermometer (Lake Shore Cryotronics, Inc.) and Hall sensors. Thermal conductivity $\kappa(T)$, specific heat $C(T)$ and charge transport characteristics were studied also using a commercial installation PPMS-9 (Quantum Design Inc.) in the range 1.9-400 K with the temperature gradient $\Delta\mathbf{T}$//[110].

The measurements of both resistivity and Seebeck coefficient in the temperature range 300- 900K were carried out using the commercial setup (Cryotel Ltd). Seebeck coefficient was detected by measuring the temperature gradient between thermocouples (Type S) pressed to the surface of the studied rectangular samples against their long sides (~12-15 mm), followed by measurement of thermal electromotive force $dE$ in the same points using the Pt wire of the thermocouples. Temperature gradients were up to 2K/mm depending on temperature. To check the correctness of experimental results obtained we used two measurement channels with distance between thermocouples 3 mm and 9 mm, respectively. High-temperature measurements of resistivity were performed by the DC four-probe technique. A direct current was applied between the ends of the sample and voltage $dV$ between the potential terminals (the Pt wires of the thermocouples) was recorded. The obtained values were corrected taking into account the contribution of thermoelectric power. All measurements were carried out in the helium gas atmosphere.

The single crystal required for X-ray diffraction (XRD) data collection was cut from the same ingot as samples #1 - #4. Spherical samples of about 0.3 mm in diameter were prepared for XRD data collection (see [40, 61] for more detail). The XRD data sets (Mo$K_\alpha$ radiation, $\lambda = 0.70930$ Å) were obtained at 27 temperatures in the temperature range 30 – 450 K using an Xcalibur and Rigaku XtaLAB Synergy-DW (Oxford Diffraction) diffractometers. To cool the sample, a Cobra Plus (Oxford Cryosystems) and N-Helix 800 Series cryosystems were applied, correspondingly, with an open stream of cold nitrogen/helium directed at the sample. The stability of temperature maintenance is ~0.1 K. Due to the temperature gradient between the sample and the temperature sensor fixed in the cryo-head nozzle at a distance of 6–10 mm from the sample studied, the measured temperatures were corrected based on the results of calibration [38].

# EXPERIMENTAL RESULTS

**I. Crystal structure and atomic vibrations.** The crystal structure of ZrB$_{12}$, which is typical for most dodecaborides, is schematically shown in Fig. 1. The structure of ZrB$_{12}$ was refined at 27 temperatures using Jana2006 program [62] in the $Fm\bar{3}m$ group of symmetry. The $F$-centered unit cell with $a_{cub} \approx 7.4$ Å contains four formula units ZrB$_{12}$ but only two atoms belong to a small, symmetry-independent volume of the cell, namely, Zr (0,0,0) in the fixed position 4$a$ and B (1/2, $y$, $y$), $y \approx 1/6$, in the special position 48$i$ of the $Fm\bar{3}m$ group. The atomic displacement parameters {$u_{ij}$}, abbreviated ADP, also known as Debye-Waller factors, were refined in the harmonic approximation. Temperature-dependent values of equivalent atomic displacement parameters $u_{eq} = (u_{11}+u_{22}+u_{33})/3$ were approximated using the Einstein and Debye models with the characteristic Einstein ($\Theta_E$) and Debye ($\Theta_D$) temperatures calculated by DebyeFit program (see [63, 41] for more detail). As can be seen from Fig. 2, the amplitudes of the root-mean-square displacements $u_{eq}$ of boron atoms in ZrB$_{12}$ are noticeably smaller than in LuB$_{12}$, whereas their values for Zr exceed those obtained for Lu. Near $T \sim 100$ K the step-like anomalies are observed in the experimental dependences $u_{eq}$(Zr) and $u_{eq}$(B) in ZrB$_{12}$, and above and below 100 K the ADPs were approximated by two Einstein and two Debye curves, correspondingly. The low-temperature and high-temperature Einstein curves allow estimating values $\Theta_E = 227$ K and $\Theta_E = 182$ K, correspondingly. Note that $\Theta_E$(ZrB$_{12}$) $\approx 200$ K was obtained by Sluchanko et al. [24] from the specific heat measurements and related closely to the energy of dispersion-less branch in the phonon spectra of ZrB$_{12}$ detected at ~17.5 meV in the inelastic neutron scattering experiments [11]. The ADP values $u_{eq}$(Zr) and $u_{eq}$(B) in ZrB$_{12}$ become equal near $\Theta_E$(ZrB$_{12}$) = 182 K (Fig. 2), whereas for LuB$_{12}$ the temperature curves $u_{eq}(T)$ of lutetium and boron do not intersect in the studied temperature range, and the Einstein temperature of the Lu ion $\Theta_E$(LuB$_{12}$) = 162(1) K is noticeably lower. The value coincides with good accuracy with the position of the dispersion-less branch in the phonon spectra of LuB$_{12}$ observed at energies of ~15 meV [17, 11, 12] and with the value of the Einstein temperature $\Theta_E$(LuB$_{12}$)= 165 K found in [41] from heat capacity measurements.

At low temperatures the experimental dependence of the boron ADP $u_{eq}$(B) in ZrB$_{12}$ is approximated by the Debye curve with $\Theta_D = 1240$ K. The break in the curve above 100 K is explained by the abrupt increase in the static component of the atomic displacement parameters $u_{eq}$(B), and in the range $T>100$ K the $\Theta_D = 1080$ K is detected both for ZrB$_{12}$ and LuB$_{12}$ in the present analysis. For comparison, value $\Theta_D = 1040$ K was obtained from the ultrasonic measurements in ZrB$_{12}$ [64] and $\Theta_D$(LuB$_{12}$) = 1077(18) K was estimated for LuB$_{12}$ in the XRD experiments [41].

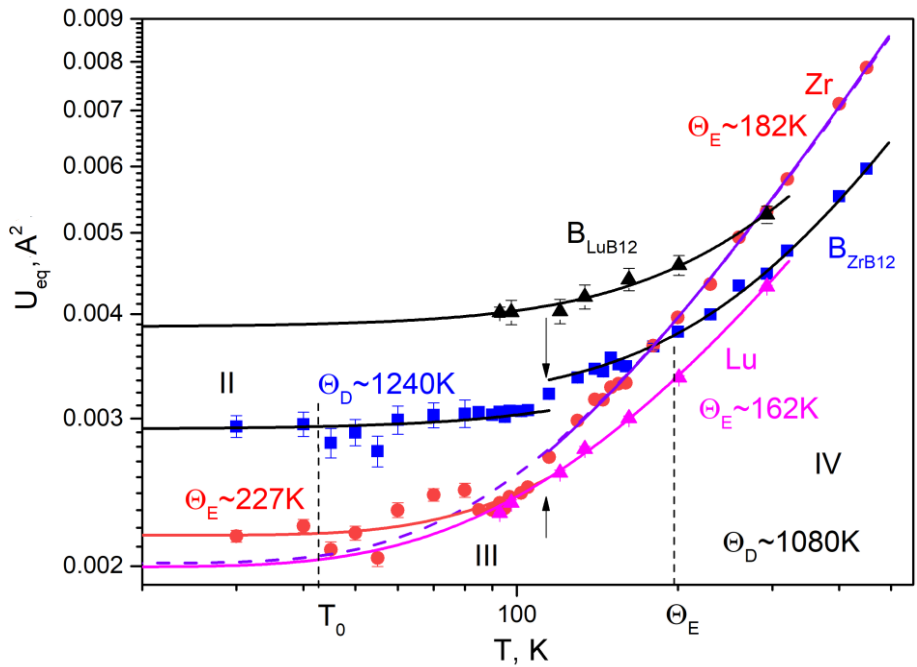

Fig. 2. Temperature dependences of $u_{eq}$ for Zr and B in ZrB$_{12}$ and Lu and B in LuB$_{12}$, approximated by Einstein and Debye models for the metal and boron atoms, respectively. Circles, squares and triangles indicate the $u_{eq}$ values determined from the XRD data sets, solid lines are the results of fitting. Roman numerals II-IV correspond to temperature intervals with different phases and regimes in ZrB$_{12}$ separated by vertical dotted lines.

The unit cell parameters and angles are distorted by the static Jahn-Teller effect being determined from the X-ray data without imposition of symmetrical bonds (Fig. 3), but distortions are very small and do not require a transition to a less symmetric model to refine the atomic coordinates and the atomic displacement parameters. The instability of the lattice parameters (Fig. 3) in the range of 80-180 K remains in the cubic model, as one can see on the temperature dependences of the interatomic distances r(Zr – B) (Fig. 4a) and the bond lengths r(B – B)$_{intra}$ in the B$_{12}$ clusters and between them r(B – B)$_{inter}$ (Figs. 4b-4c). At temperatures below 80 K, the parameters r(Zr – B), r(B – B)$_{intra}$ and r(B – B)$_{inter}$ are virtually independent of temperature.

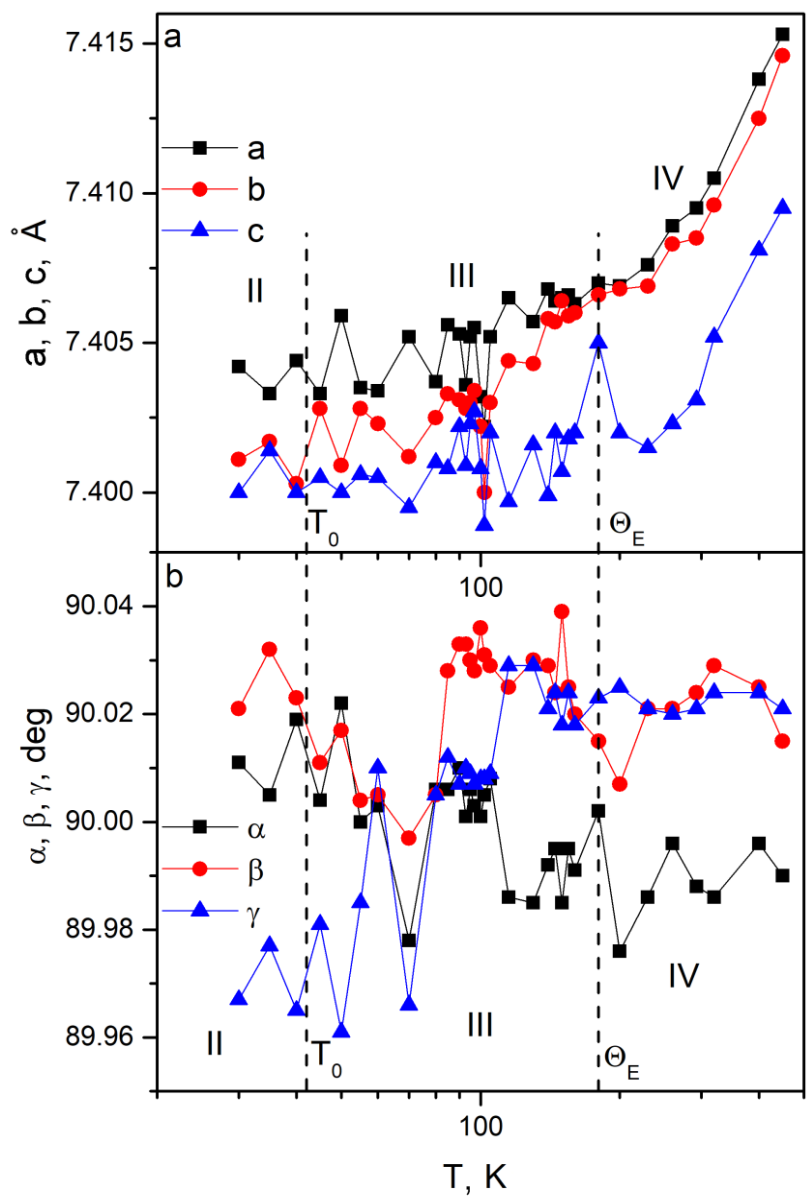

Fig. 3. Temperature dependences of (a) linear and (b) angular parameters of the unit-cell of ZrB$_{12}$, determined from XRD data sets without imposing symmetry restrictions. Roman numerals II-IV correspond to temperature intervals with different phases and regimes in ZrB$_{12}$ separated by vertical dotted lines.

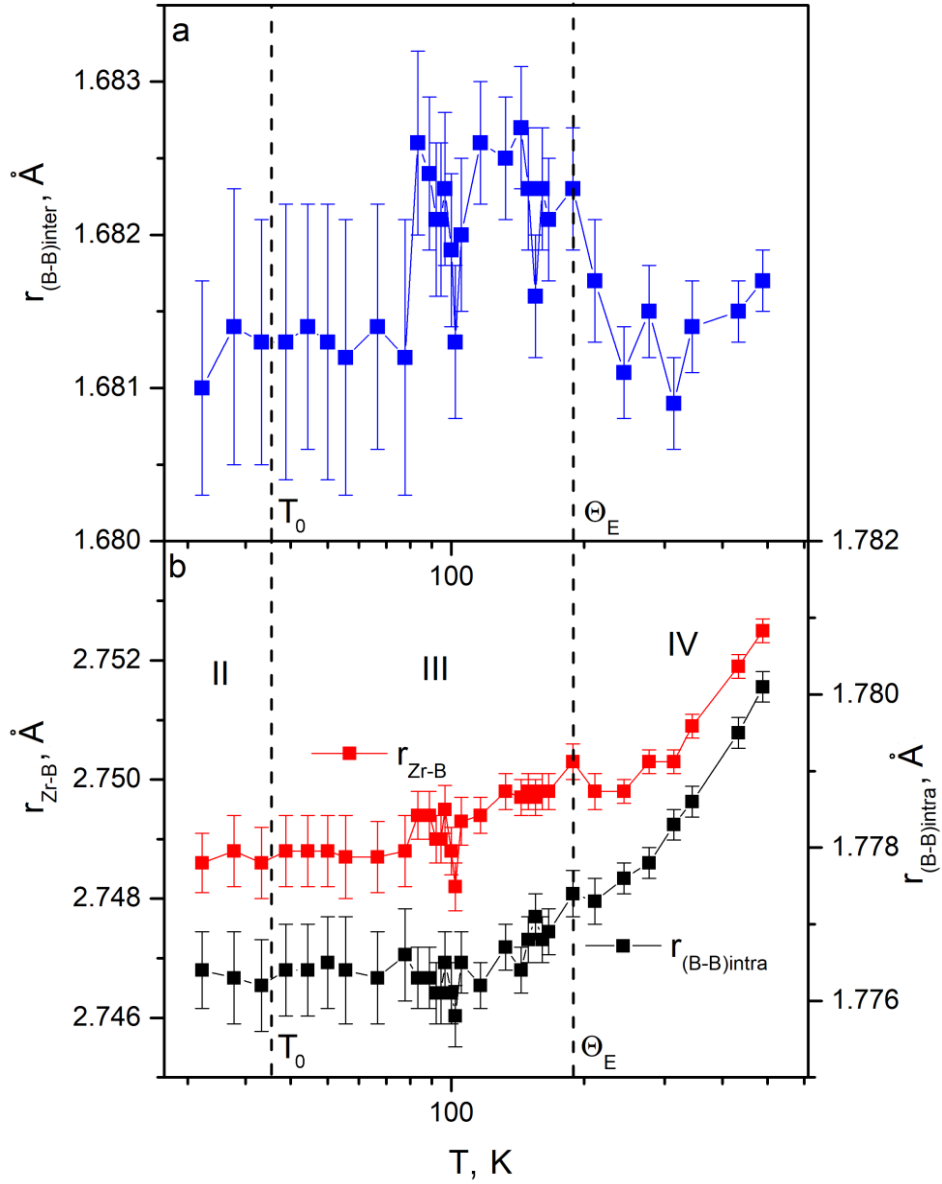

Fig. 4. Temperature-dependent distances (a) r(Zr – B) and (b) r(B – B)$_{intra}$ between the boron atoms in one cuboctahedron and r(B – B)$_{inter}$ between the nearest boron atoms in neighboring cuboctahedra B$_{12}$ in ZrB$_{12}$. Roman numerals II-IV correspond to temperature intervals with different phases and regimes in ZrB$_{12}$ separated by vertical dotted lines.

**II. s-CDW and dynamic charge stripes.** As shown in [53], in ZrB$_{12}$, in contrast to LuB$_{12}$, at temperatures below 100 K, antinodes of the sub-structural charge density wave (s-CDW) with a period approximately equal to the distances between boron atoms r(B – B)~1.7 Å are registered in the interstices of the crystal lattice. Fig. 5 shows the distribution of electron density (ED) obtained by the maximum entropy method (MEM) in the range of 30–293 K, in the (111) planes passing through the triangular faces of the B$_{12}$ cuboctahedra in ZrB$_{12}$ (see Fig. 1).The distribution pattern indicates a noticeable presence of s-CDW in the temperature range studied. Note that in LuB$_{12}$ and other rare earth dodecaborides with a lower concentration of charge carriers in the conduction band compared to ZrB$_{12}$, the antinodes in the indicated temperature range are less pronounced, or not observed, see, for example, Fig. S1 in the Supplementary Materials.

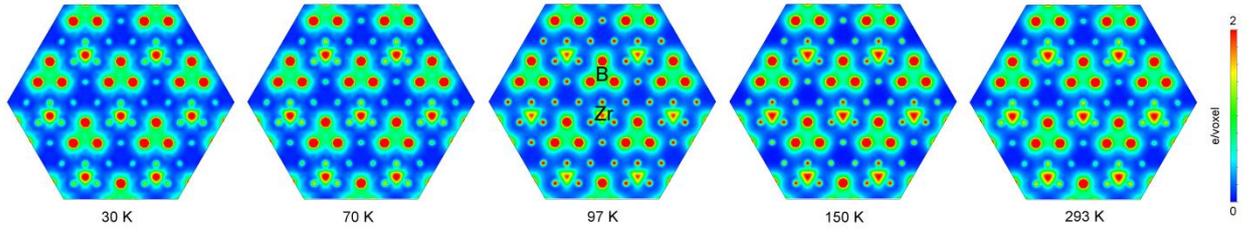

Fig. 5. MEM maps of the ED distribution in the (111) plane of ZrB$_{12}$ at room and low temperatures. The calculations carried out taking into account the cubic symmetry of the structural model. Large red circles are boron atoms in the triangular faces of B$_{12}$ cuboctahedra. Zr atoms, one of which is marked on the central map, are out of plane by ~0.7 Å. The ED peaks are truncated at a height of 2 e/voxel to show in color the fine details of the ED distribution. Antinodes of the sub-structural charge density wave (s-CDW) are observed in the interstices of the crystal lattice with a period approximately equal to the distances between boron atoms, r(B – B)~1.7 Å.

At the same time, in the planes of the {110} family passing through the Zr ions, with a decrease in temperature in the range of 30-70 K, dynamic charge stripes are formed in the [110] directions, localized in the chains of boron atoms (Fig. 6) and caused by the dynamic JT instability of the boron framework. The maximum values of ED in stripes are between B$_{12}$ clusters and near the surface of B$_{12}$ clusters. The distribution of ED in the (100) planes passing through the Zr ions and the square faces of the B$_{12}$ cuboctahedra is shown in Fig. 7a (see also Fig. 1) and Fig. 7b, respectively. As can be seen from Figs. 6-7, near Zr the ED distribution is deformed from spherical, which indicates the formation of vibrationally coupled Zr-Zr pairs oriented both in the {100} directions to the nearest B$_{12}$ clusters and along {112} in the direction of the bond between B$_{12}$ clusters (two Zr-Zr pairs are indicated by arrows in Fig. 6). Thus, it should be assumed that the dynamic charge stripes in the chains along {110} in ZrB$_{12}$ include not only the 2$p$ states of the boron framework but also the 4$d$ conduction electrons supplied by Zr$^{4+}$ ions. In contrast to ZrB$_{12}$ at low temperatures, where the dynamic charge stripes consist predominantly of 2$p$ electrons of boron and are grouped along directions {110} in the boron sub-lattice (see Figs. 15a and 15c below), in the RE dodecaborides $R$B$_{12}$ ($R$ = Dy, Ho, Er, Tm, Lu) the 5$d$-2$p$ mixed ED is observed in the stripes, located strictly along one direction in the family {110} and passed through the RE ions and the B-B bridges in the spaces between B$_{12}$ clusters (see Fig. 15e below and Fig. S2 in the Supplementary Material).

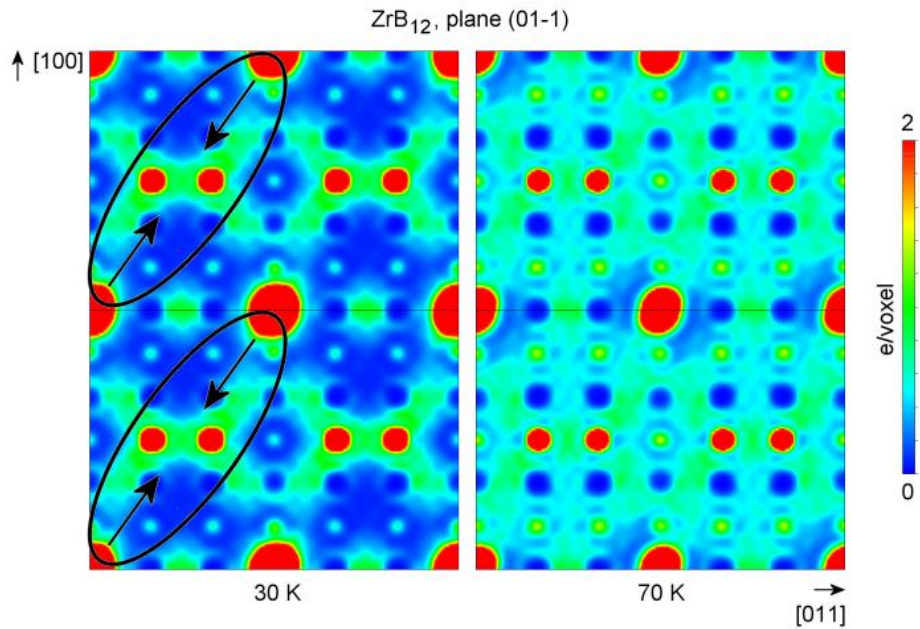

Fig. 6. MEM maps of the ED distribution in the (01-1) plane of ZrB$_{12}$ at 30 K and 70 K. The calculations were carried out without taking into account the cubic symmetry of the structural model. The ED peaks are truncated at a height of 2 e/voxel to show in color the fine details of the ED distribution in the interstices of the crystal lattice. The plane contains Zr atoms (large red circles) and B atoms (small red circles). Two Zr-Zr vibrational pairs formed along the <112> family axes are highlighted.

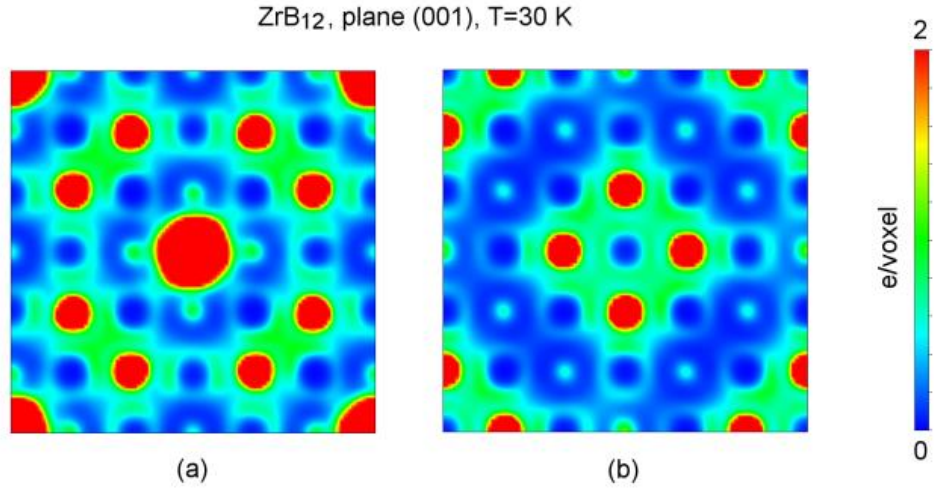

Fig. 7. MEM maps of the ED distribution in parallel sections (001) of $ZrB_{12}$ passing through Zr ions (a) and square faces of $B_{12}$ cuboctahedra (b). The calculations carried out without taking into account the cubic symmetry of the structural model. The ED peaks are truncated at a height of 2 e/voxel to show in color the fine details of the ED distribution in the interstices of the crystal lattice.

### III. Physical properties.

**III.1. Resistivity.** Fig. 8 shows the temperature dependences of resistivity $\rho(T)$ in the range 1.8-300 K of several $ZrB_{12}$ single crystals with different DC orientation ($\boldsymbol{J} \parallel$ <100> in samples #1-#3 and $\boldsymbol{J} \parallel$ [110] in #4) and $LuB_{12}$ for comparison. The $\rho(T)$ curves of $ZrB_{12}$ exhibit a typical metallic behavior with a practically linear temperature dependence above $T \sim 150$ K and a rather small residual resistivity ratio $\rho(300\ K)/\rho_0 = 7.8–8.3$. The large-scale plot in the inset to Fig. 8 demonstrates the superconducting transition with a small width $\Delta T_c^{(\rho)} \sim 0.05$ K for all the samples studied. The superconductivity in ZrB$_{12}$ with $T_c \approx 5.95$ K suppressed completely by strong magnetic field (inset to Fig. 8). It is worth noting that both the room temperature resistivity values (~19 µOhm·cm, Fig.8) and the slope of about linear $\rho(T)$ dependence are practically the same for all samples with different DC orientation, and the residual resistivity $\rho_0$ changes moderately (2.35-2.65 µOhm·cm, see inset to Fig. 8) in good agreement with previously reported values [14]. The results contradict to very strong $\rho(T)$ anisotropy, which was deduced for ZrB$_{12}$ crystals in [65], where a difference of about 3 times was observed both for the room temperature resistivity and the values of the residual resistivity $\rho_0$ measured for ***J*** ∥ <100> and ***J*** ∥ <110> DC orientation. The authors [65] proposed that such a strong anisotropy may be explained by the small amount of needle-shaped inclusions of ZrB$_2$ diboride erected in the ZrB$_{12}$ samples.

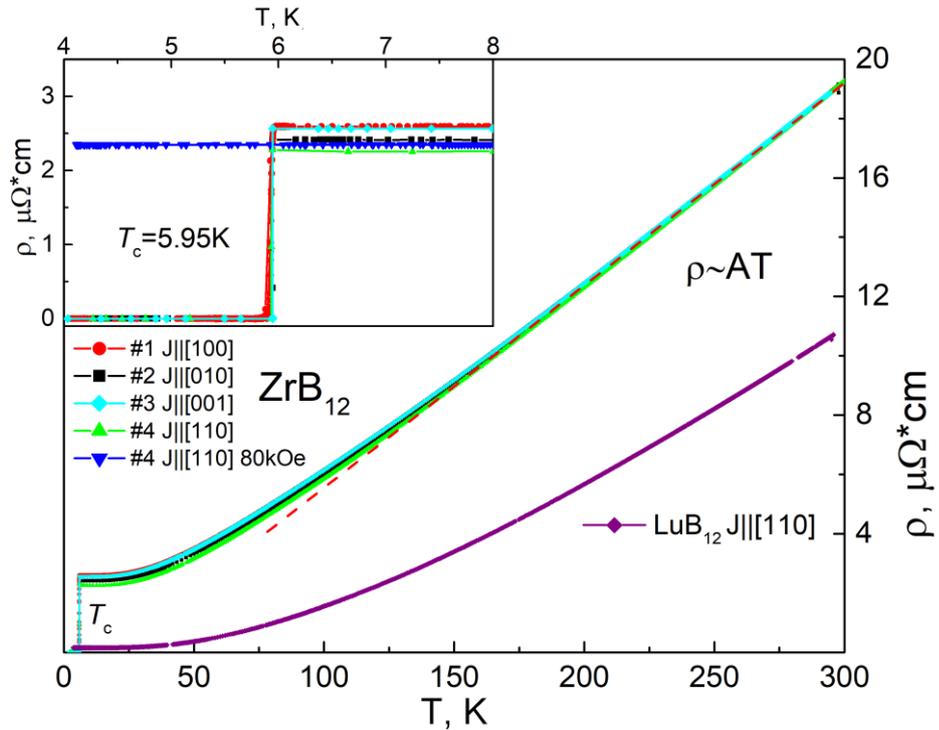

Fig. 8. Resistivity temperature dependences for $ZrB_{12}$ samples. Dashed line demonstrates the linear approximation. Inset shows the large-scale view of the superconducting transition. For comparison, the $\rho(T)$ curve for $LuB_{12}$ is shown.

Taking into account the loosely bound state of Zr/Lu-ions in the *fcc* UB$_{12}$-type crystal structure (Fig. 1), it is natural to expect that the resistivity temperature variation could be approximated by a sum of two terms which arise additionally to residual resistivity as it was detected previously for LuB$_{12}$ [41]. Among these, the low temperature Einstein component $\rho_E$ [66]

$$\Delta\rho = \rho - \rho_0 = \rho_E = \frac{A}{T\left(e^{\frac{\theta_E}{T}}-1\right)\left(1-e^{-\frac{\theta_E}{T}}\right)} \tag{1a}$$

was observed for LuB$_{12}$ in [41] in the range $T < T^* \sim 60$ K in the cage-glass state [23] (see Fig. 9a) and it was attributed to the strong electron-phonon scattering on quasi-local vibrations (Einstein modes) of the Lu$^{3+}$-ions [41]. In contrast to the result obtained for LuB$_{12}$, the Fermi-liquid behavior $\Delta\rho \sim T^2$ was found for ZrB$_{12}$ at low temperatures $T<T_0 \approx 42$ K (Fig. 9a), which is not in accord with previously reported power-law dependence $\Delta\rho \sim T^5$ [14]. The second resistivity term is detected above 100 K both in LuB$_{12}$ [41] and ZrB$_{12}$ (see Fig. 9a) and this is due to the Umklapp processes dominating at least in LuB$_{12}$ at high temperatures. The resistivity variation was approximated by the relation [41]

$$\Delta\rho = \rho_U = B_0 T e^{-\frac{T_0}{T}} \tag{1b}$$

(shown by blue lines in Fig. 9a). For highly accurate analysis of $\Delta\rho(T)$ components, resistivity measurements were carried out in ZrB$_{12}$ and LuB$_{12}$ in a wide temperature range of 77 – 900 K (Fig. 9a). For LuB$_{12}$, it can be seen that the temperature dependence of resistivity is very well described by Eq. (1b) at intermediate and high temperatures, and parameter $T_0 = 171.2$ K is close to the Einstein temperature $\Theta_E = 162$ K. On the contrary, for ZrB$_{12}$ the deduced value $T_0 \approx 42$ K is quite different from the $\Theta_E = 182$ K (Fig. 2). The discrepancy will be discussed in more detail below.

**III.2. Seebeck coefficient.** Wide-range temperature dependences of Seebeck coefficient of ZrB$_{12}$ and LuB$_{12}$ are shown in Fig. 9b. The $S(T)$ curves are also quite different in the studied dodecaborides, although demonstrating distinct step-like anomalies below 200 K. Indeed, in the case of LuB$_{12}$ the Einstein relation for the thermopower

$$S_E(T) = S_{E0}\left(\frac{\theta_E}{T}\right)^2 \frac{e^{\frac{\theta_E}{T}}}{\left(e^{\frac{\theta_E}{T}}-1\right)^2} \tag{2a}$$

gives a good approximation in the range 10-200 K for the temperature gradient ***ΔT***//[110] with $\Theta_E = 162$ K, and the tendency to power-law dependence is observed above 400 K (Fig. 9b). Note, that the step-like Einstein anomaly in LuB$_{12}$ is independent of magnetic field above 30K, but there is a very small field-induced positive maximum ($\Delta S(T) < 1$ µV/K, Fig. 9b), which is observed in the range 10-30 K. On the contrary, for ZrB$_{12}$ the step-like singularity of Seebeck coefficient is well described by relation

$$S(T) = S_0 e^{-\frac{T_0}{T}} \tag{2b}$$

with the same value $T_0 \approx 42$ K. It is worth noting that the step-like anomaly in ZrB$_{12}$ depends significantly on the magnetic field, and at $H=90$ kOe we get $T_0 \approx 52$ K with the same pre-exponential $S_0 \approx 4.2$ µV/K (Fig.

9b). The zeroing of the Seebeck coefficient, found here for ZrB$_{12}$ below 20 K and well above the superconducting transition at $T_c \approx$ 6 K, is usually a sign of a sliding charge density wave [67]. We propose that the parameter $T_0$ in Eq. (2b) corresponds to the CDW gap value $\Delta_0 \approx$ 42 K, which increases up to 52 K in magnetic field, and the gap widening attributed likely with field-induced ED redistribution from dynamic charge stripes to s-CDW. A similar field-induced effect was discovered for NdSe$_3$ [68], where an increase in the CDW was observed in a magnetic field of up to 300 kOe, accompanied by an approximately linear increase in $T_0$ in the range of 59–80 K.

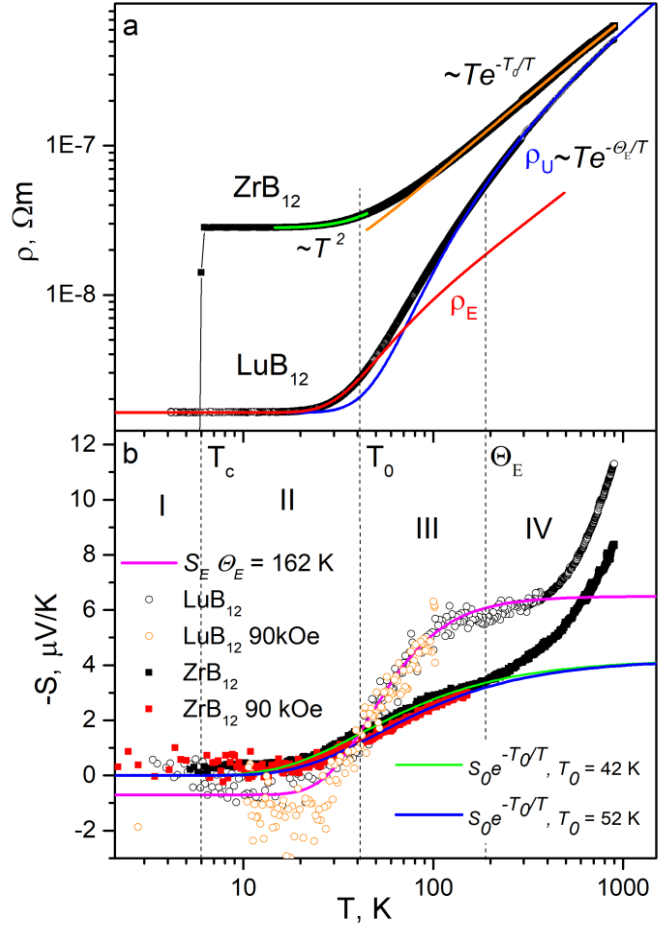

Fig. 9. Temperature dependences of (a) resistivity $\rho(T)$ and (b) Seebeck coefficient $S(T)$ (temperature gradient $\Delta T$//$H$//[110]) for ZrB$_{12}$ and LuB$_{12}$ in the wide temperature range 2-900 K. The approximations by the Einstein model Eqs. (1a), (2a), $\rho_U(T)$ by Eq.(1b), and the power-law $\rho \sim T^2$ are shown by the solid lines. In panel (b) the approximation by Einstein relation (Eq.(2a) for LuB$_{12}$ and two fits for ZrB$_{12}$ by Eq.(2b) at $H$=0 and 90 kOe are shown by solid lines, see text). Roman numerals I-IV correspond to temperature intervals with different phases and regimes in ZrB$_{12}$.

**III.3. Thermal conductivity**. Thermal conductivity $\kappa(T)$ of LuB$_{12}$ reaches a maximum in zero field at a temperature of about 25 K (see also [69]) and near 30 K in a strong magnetic field $H$=90 kOe (Fig. 10a), and the almost linear low temperature thermal conductivity of LuB$_{12}$ decreases by 15-25% in the external magnetic field. The unusual response to magnetic field in the non-magnetic metal LuB$_{12}$ with linear dynamic charge stripes ([43], and Fig. 15e below) may be due to the effects of field-induced reorientation of these inhomogeneous structures of fluctuating charges, which leads to a noticeable dissipation of the heat flow. Anomalous low-temperature components caused by stripes are observed in LuB$_{12}$ in both the Hall effect and magnetoresistance [44], strongly weakening as the temperature increases above 25–30 K. In addition, a step-like anomaly in the strong-field NMR line width of $^{175}$Lu was detected at $T$~30 K and attributed to the crystallization of "structural defects" in the LuB$_{12}$ matrix [70].

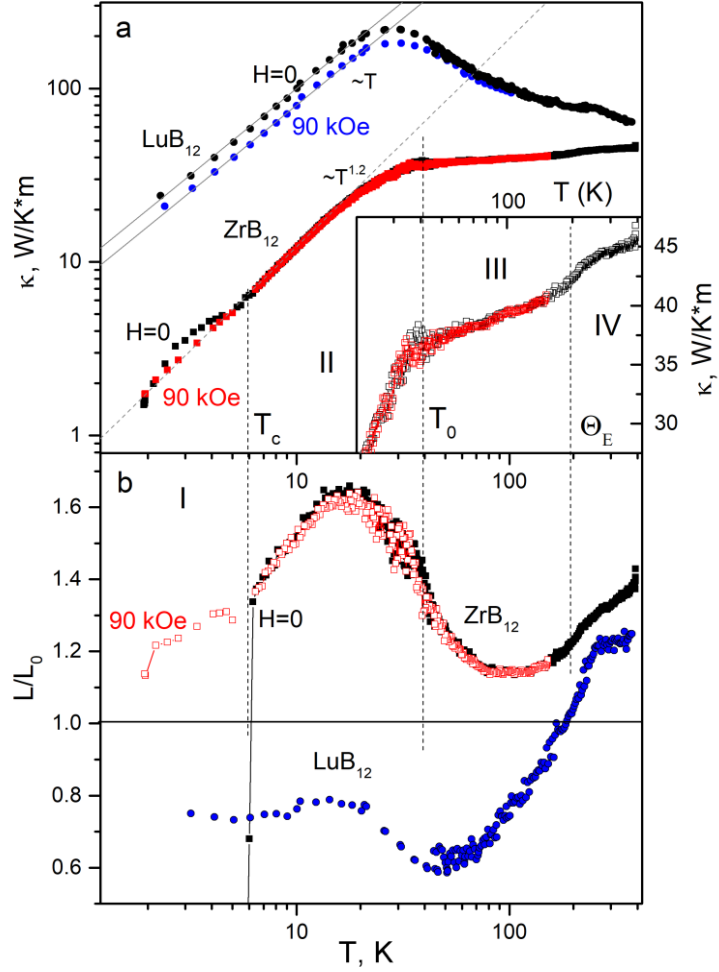

Fig. 10. Temperature dependences of (a) thermal conductivity $\kappa(T, H)$ LuB$_{12}$ and ZrB$_{12}$ at $H$=0 and 90 kOe and (b) normalized Lorentz number $L/L_0(T)$ (see text). The inset to panel (a) shows the large scale plot of $\kappa(T, H)$ in ZrB$_{12}$ in the range 20-400 K. Roman numerals I-IV correspond to temperature intervals with different phases and regimes in ZrB$_{12}$.

Unlike LuB$_{12}$, $\kappa(T)$ dependence in ZrB$_{12}$ demonstrates continuous decrease with the temperature lowering with three singularities at (*i*) $T_c$~ 6 K, (*ii*) $T_0$~42 K and (*iii*) $T$~$\Theta_E$~195 K (Fig. 10a). Complete suppression of superconductivity below $T_c$ (interval I) in the magnetic field $H$=90 kOe is accompanied by the restoration of a practically linear dependence $\kappa(T)$ in the entire range of 2–20 K, and the only moderate field-induced decrease is observed near the phase transition at $T_0 \approx 42$ K (see inset to Fig. 10a). Fig. 10b presents the normalized Lorentz number $L/L_0(T)$ calculated using the Wiedemann-Franz relation $\kappa = L_0 T/\rho$ (where $L_0$ = 24.5 nW·Ω·K$^{-2}$ is the Sommerfeld value of Lorentz number for free electron gas). It can be seen that the significant difference in $L/L_0$ between LuB$_{12}$ and ZrB$_{12}$ appears below 180 K (interval III), where strong s-CDW antinodes arise on the MEM maps of electron density (Fig. 5). We suppose that the transition at $T$~$\Theta_E$~180 K to the sliding CDW state should be responsible for the $L/L_0$ increase of $L/L_0$ in ZrB$_{12}$, and below $T_0$~42 K (interval II) the s-CDWs are pinned to the crystal inhomogeneities.

**III.4. Heat capacity.** The temperature dependences of the specific heat $C(T)$ of ZrB$_{12}$ and LuB$_{12}$ are shown in Fig. 11a. As can be seen, the gradual decrease of heat capacity at temperatures from 300 K to 40 K is accompanied by a sharp and almost step-like decrease, which is a typical Einstein-type $C(T)$ dependence. It is worth noting that although the $C(T)$ curves of ZrB$_{12}$ and LuB$_{12}$ samples at $T$ >10 K are almost identical in the double logarithmic plot used in Fig. 11a, the position of the step-like anomaly is shifted upward along the $T$ axis in ZrB$_{12}$ with $\Theta_E \approx 195$ K compared to LuB$_{12}$ ($\Theta_E \approx 162$ K), and in ZrB$_{12}$ the singularity becomes broader. Fig. 11a shows also derivative $dC/dT=f(T)$, which demonstrate clearly the position of the step-like Einstein anomaly of $C(T)$ in ZrB$_{12}$. The analysis of the specific heat with separation of $C(T)$ contributions was carried out for LuB$_{12}$ in [41], where it has been shown that Debye ($C_D$) and

Einstein ($C_E$) components dominate at intermediate temperatures. In the case of ZrB$_{12}$, there are three distinct features of heat capacity detected here, including (*i*) a jump at the superconducting transition temperature at $T_c\sim 6$ K, (*ii*) a knee with strong instability of $dC/dT(T)$ near $T_0\sim 42$ K and (*iii*) a gentle maximum of $dC/dT$ near $\Theta_E$ (Fig. 11a). The last two singularities at $T_0$ and $\Theta_E$ correlate very well with the transitions in the s-CDW state in ZrB$_{12}$, which is expected to be accompanied with additional contributions to the heat capacity [71]. To study the $C(T)$ anomalies near $T_0$ and $\Theta_E$ we plotted and analyzed the $C_D(T)$, $C_E(T)$ and the linear Sommerfeld components in Fig. 11b, estimating the difference $\Delta C(T)=C-C_D-C_E-\gamma T$, where Sommerfeld coefficient $\gamma = 4.6$ mJ/(mol K$^2$), $\Theta_E=195$ K and $\Theta_D=1260$ K were used. In Fig. 11b it is clearly seen that there are two maxima of $\Delta C(T)$ near $T_0$ and $\Theta_E$, marked by vertical dotted lines, which probably correspond to hidden phase transitions in ZrB$_{12}$.

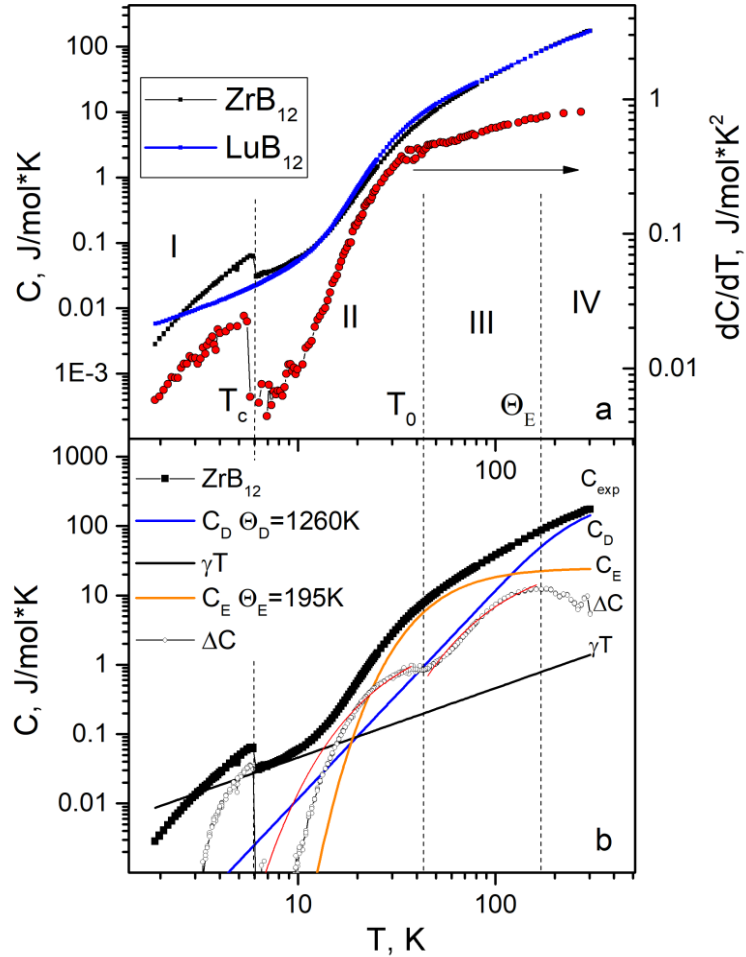

Fig. 11. (a) Temperature dependences of specific heat $C(T)$ for LuB$_{12}$ and ZrB$_{12}$ and the derivative $dC/dT=f(T)$ for ZrB$_{12}$. Panel (b) shows for ZrB$_{12}$ the Debye ($C_D$), Einstein ($C_E$), linear Sommerfeld $\gamma T$ contributions to $C(T)$ and the difference $\Delta C(T)$ that demonstrates two phase transitions near $T_0$ and $\Theta_E$ (see text). Roman numerals I-IV correspond to temperature intervals with different phases and regimes in ZrB$_{12}$.

**III.5. Magnetoresistance and Hall effect.** Field and angular dependences of positive magnetoresistance (MR) $\Delta\rho/\rho(H, \varphi)$ were studied here for the samples #1-#4 of ZrB$_{12}$ at liquid helium temperature. Additionally, for the same samples #1-#4 we investigated the high-field temperature and angular dependences of negative Hall coefficient $R_H(T, \varphi, H=80$ kOe$)=\rho_H/H$ ($\rho_H$ is the Hall resistivity). Figs.12-13 show the field and angular MR curves, correspondingly, recorded for various **H-J** configurations. For comparison, the MR dependences of LuB$_{12}$ are also presented in Fig. 12 and Fig. 13b. It is clearly seen from Figs.12-13 that the MR of ZrB$_{12}$ is practically isotropic (the anisotropy does not exceed 0.25%, see Fig.13b) and the amplitude of magnetoresistance does not exceed 4% in the magnetic field up to 80 kOe. The result is quite different from the MR of LuB$_{12}$ [29, 40, 32, 72, 44], where the $\Delta\rho/\rho$ ~500-800% has been established (see, for example, curve for LuB$_{12}$ in Fig. 12), and a huge MR anisotropy was detected (~300%, see, for example, Fig. 13b). For ZrB$_{12}$, an almost quadratic behavior of the MR $\Delta\rho/\rho \approx \mu_D^2 H^2$ was found in the range of 50-80 kOe (Fig. 12), and a rather low drift mobility of charge carriers was estimated $\mu_D \approx 200\text{-}220$ cm²/(V s), which attributes all ZrB$_{12}$ samples to a weak-field regime ($\omega_c\tau \ll 1$, where $\omega_c$ is the cyclotron frequency) of charge transport with strong scattering of carriers.

We propose that the dramatically different charge transport anisotropy of LuB$_{12}$ and ZrB$_{12}$ (Figs. 12-13) can be attributed to the charge carrier scattering on the dynamic charge stripes, which were found to be close to linear and located along one of the <110> directions in LuB$_{12}$ [40], see Fig.15e below as an example. In contrast, ZrB$_{12}$ exhibits two types of conduction electron density singularities, in the form of (*i*) 3D grids of dynamic charge stripes along the <110> directions (Fig. 6, left panel) and (*ii*) a triangular lattice of s-CDW antinodes located at the boron lattice interstices in the {111} planes (Figs. 5 and 6, left panel). As a result, we propose a quantum diffusion regime of charge transport at temperatures above $T_0$(ZrB$_{12}$)≈ 42 K with a small mean free path $l$ < 130 Å of charge carriers [48] and a nearly linear temperature dependence of the resistivity (Fig. 9a).

The Hall coefficient was measured for samples #1-#3 of ZrB$_{12}$ with three orientations of the measuring current $J \parallel [100]$, $J \parallel [010]$ and $J \parallel [001]$ both (*i*) by the field-sweep measurements in two opposite directions of the external magnetic field $\pm H \parallel n \parallel$ <001> (*n* is the normal vector to the lateral surface), and (*ii*) recording the angular dependences of the Hall resistivity for $n \parallel$ <001> in an experiment with sample rotation (see the scheme in the inset to Fig. 12 and an example of $\rho_H/H(\varphi)$ curves in Fig. 14b). In the case of DC configuration $J \parallel [110]$ (the sample #4) two different orientations $H \parallel n \parallel [001]$ and $H \parallel n \parallel [011]$ were used. A collection of temperature dependences measured at $H$=80 kOe is presented in Fig. 14a, which allows us to compare the experimental results $R_H=\rho_H/H$ for different field-current configurations. Error bars allow concluding opposite noticeable anisotropy of the Hall coefficient both in the coherent state (intervals I-II) and well above $T_0 \approx 42$ K (III-IV). The result is quite different from the Hall effect detected in LuB$_{12}$, where a positive stripe-induced component of $R_H$(T) was found to be the same amplitude as the ordinary negative Hall coefficient ([44], see also $\rho_H/H=f(\varphi)$ curve for LuB$_{12}$ presented for comparison in Fig. 14b). It is worth noting that for any field-current configuration of ZrB$_{12}$ both of the s-CDW transitions at $T \sim \Theta_E \sim 195$ K and at $T_0 \sim 42$ K are accompanied with a moderate (~4-6%) changes of the Hall coefficient (Fig. 14a). Note also that the values $R_H(T) \approx 2.5\text{-}2.7 \cdot 10^{-4}$ cm³/C correspond to the normalized concentration of conduction electrons $n/n_{Zr} \sim 2.4$ in the approximation of one type of the charge carriers. Thus, both magnetoresistance (see Fig. 13) and Hall coefficient (Fig. 14) are close to isotropic in ZrB$_{12}$ metal with the network of dynamic charge stripes (see Fig. 15c below). Estimating from the data of Figs. 9a and 14 the Hall mobility of charge carriers $\mu_H=R_H/\rho$ at helium temperature, we obtained rather small values $\mu_H$=100-120 cm$^2$/(V·s) for all crystals of ZrB$_{12}$. Taking the effective mass $m^*\sim0.7\ m_0$ [73] we estimate a very short average relaxation time of charge carriers $\tau_{e-ph}\approx4\cdot10^{-14}$ s. Similar value $\tau_{e-ph}\approx1.74\cdot10^{-14}$ s (see blue point in Fig. 16b below) was deduced independently in [73] from the analysis of quantum oscillations in ZrB$_{12}$ in the range 0.07-4.2 K. In addition, the temperature dependence of the mobility $\mu_H\sim T^{-1}$ turns out to be valid, at least in the range $T > T_0$ (intervals III-IV), which is a consequence of the linear behavior of the resistivity (Fig. 9a).

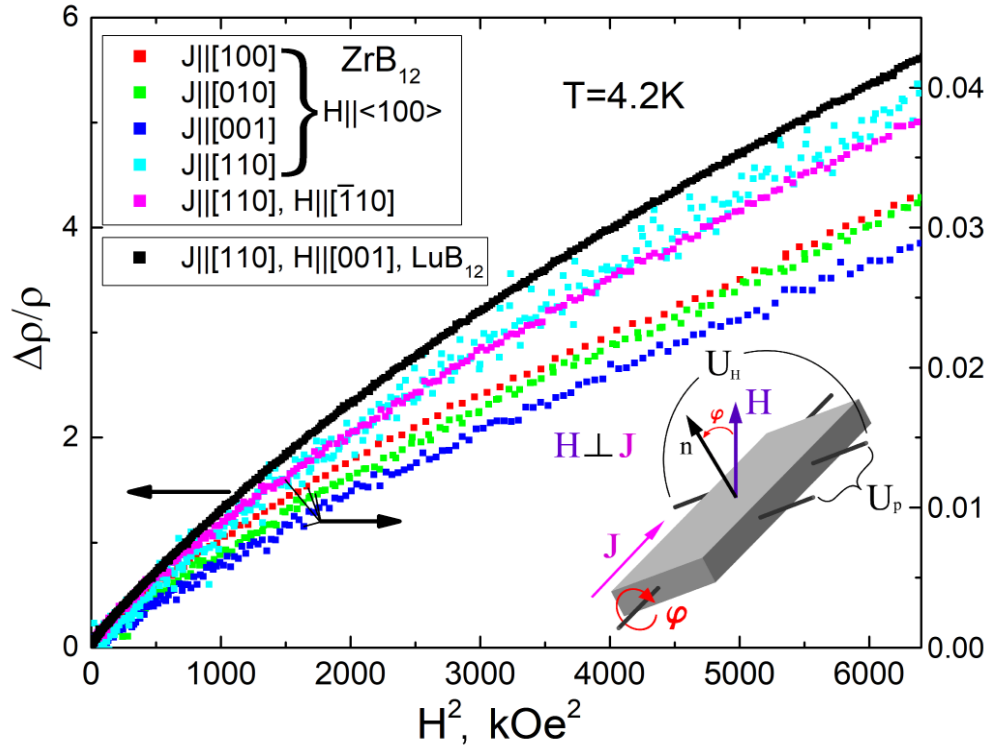

Fig. 12. Field dependences of magnetoresistance for samples #1-#4 of ZrB$_{12}$ at liquid helium temperature. The MR curve of LuB$_{12}$ is shown for comparison. Inset presents the schematic view of the angular experiment with the sample rotation around the DC axis, *n* is the normal vector to the lateral sample surface, $U_p$ and $U_H$ are voltages detected from the potential and Hall probes, correspondingly.

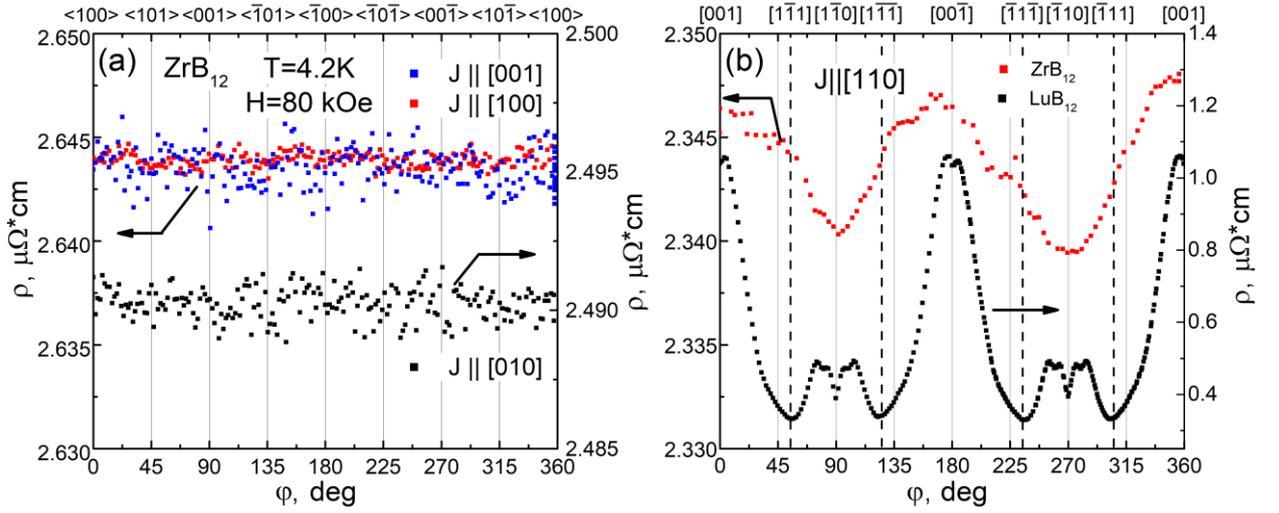

Fig. 13. Angular dependences of resistivity of ZrB$_{12}$ recorded at $H$=80 kOe for (a) the samples #1-#3 and (b) the sample #4. For comparison panel (b) demonstrates also the $\rho(\varphi)$ curve for LuB$_{12}$.

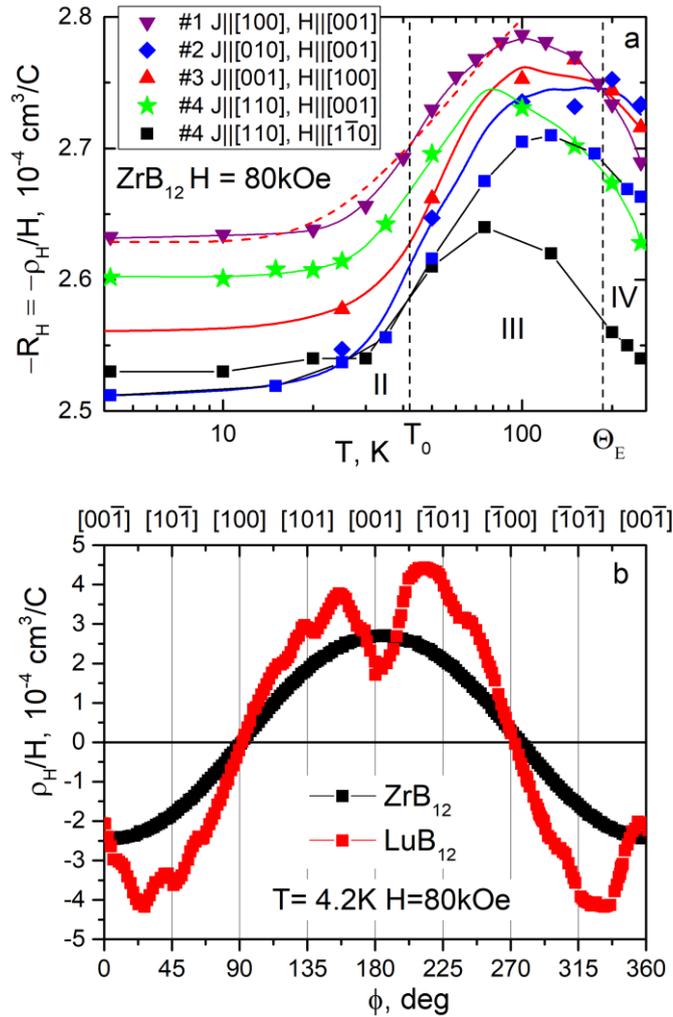

Fig. 14. (a) Temperature dependences of the Hall coefficient measured at $H$=80 kOe for the samples #1- #4 of ZrB$_{12}$ at various $\mathbf{J\text{-}H}$ configurations (see legend). Thick dashed line in panel (a) shows the data approximation by relation $R_H(T)=R_{H0}\exp(-T_0/T)$. Roman numerals II-IV correspond to temperature intervals with different phases and regimes in ZrB$_{12}$. (b) Angular dependences of the reduced Hall resistivity $\rho_H/H$ detected for the sample #1 of ZrB$_{12}$ and LuB$_{12}$ at liquid helium temperature and $H$=80 kOe. The crystallographic directions are indicated on the upper axis on panel (b).

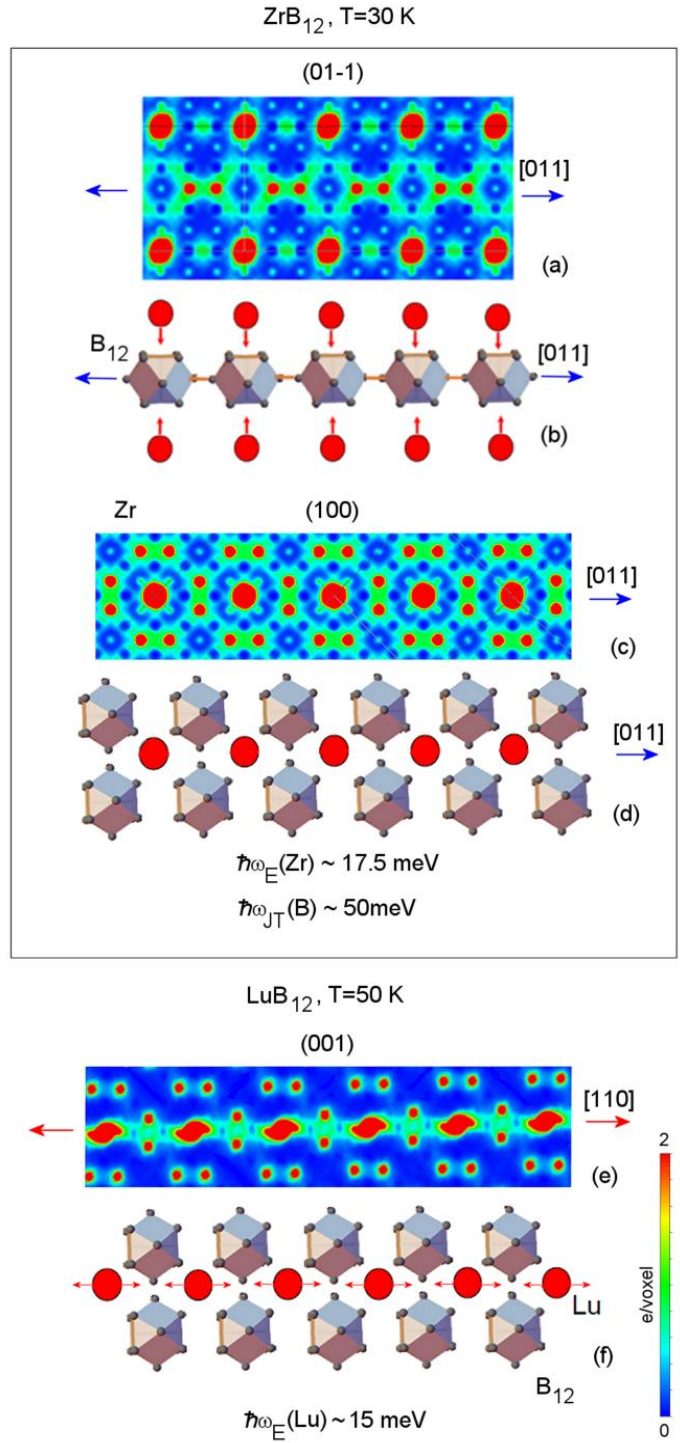

Fig. 15. MEM maps of ED distribution in (01-1) plane (a), (100) plane (c) and corresponding structural fragments (b, d) of ZrB$_{12}$ at $T$=30 K. MEM map of ED distribution in (001) plane (e) and corresponding structural fragment (f) of LuB$_{12}$ at $T$=50 K. The ED peaks are truncated at a height of 2 e/voxel to show in color the fine details of the ED distribution in the interstices of the crystal lattice.

## DISCUSSION

Let us start with a discussion of the electron density (ED) distribution in ZrB$_{12}$ and LuB$_{12}$ superconductors with very close parameters of both conduction bands and phonon spectra and with sharply different (by a factor of 15–20) superconducting transition temperatures $T_c$ and critical magnetic field $H_c$ [48]. The ED distribution determined by the maximum entropy method (MEM) in these two dodecaborides differs in two factors. The first is the sub-structural CDW in ZrB$_{12}$ (Fig. 5), which strongly changes the charge transport and thermodynamic characteristics at $T_0 \approx 42$ K and $\Theta_E$ (Figs. 9–11, 14), but was not detected in LuB$_{12}$ [53]. Note that a multiple CDW phase transitions have been observed previously in various 3D systems, and are widely believed to be related to the existence of quasi-1D chains along one crystallographic direction (see reviews [74, 75] and references therein). In some cases the competition between CDW and superconductivity develops, but both the CDW mechanism in these materials and mutual influence of these two collective states are not well understood [74, 75].

The second difference, which appears to be the most important for superconductivity in these two $R$B$_{12}$ crystals, relates to the low-temperature configuration of the dynamic charge stripes. Indeed, *5d-2p-type stripes* are arranged in LuB$_{12}$ *in linear patterns*, with the charge fluctuations, which take place on the frequencies of about 240 GHz along the unique [110] direction in the *fcc* lattice of the rare earth dodecaborides [61]. In contrast, the *network of 2p-type stripes* is observed in ZrB$_{12}$ (Fig. 6), oscillating with the frequency of Jahn-Teller collective mode of the rigid boron cage ($\hbar\omega_{JT} > 50$ meV, [76]). The JT mode induces the transverse to stripe quasi-local vibrations of Zr$^{4+}$ ions in pairs, and these Einstein phonons, which mediate the superconductivity in ZrB$_{12}$, turn out to be synchronized by the quasi-1D collective dynamics of the boron chains. Figs. 15a-d demonstrate in more detail the proposed scenario of emergence of the 1D-dynamic chains in combination with vibrationally coupled Zr-pairs in ZrB$_{12}$. We propose that the attraction between two electrons located inside the Zr-Zr pair is mediated by both the quasi-local oscillations of Zr-ions (phonons) and the high frequency collective JT-vibration in the chains (plasmons). As a result, the composite *plasmon-phonon pairing mechanism* includes two quasi-local vibrations (Einstein modes) of Zr- ions separated by the coherence length (~570 Å, [50]) and synchronized by collective 1D-oscillations of B$_{12}$ clusters (see sketches of two transverse planes with two-types pictures of stripes and Zr-ions in Figs. 15a-b, 15c-d). The low-frequency 5d-2p stripe found in LuB$_{12}$ (Fig. 15e, [40]) involves a combination of (*i*) Lu ion vibrations with (*ii*) boron pair vibrations and (*iii*) electron density fluctuations along the one-dimensional chain, so the proposed plasmon-phonon composite pairing mechanism is not valid in this case. Let us point out, that the configuration of dynamic charge stripe with the vibrationally coupled *Zr-pairs, which are transverse to stripe*, is quite similar to one long chain on which the series of lateral chains are attached, as it was proposed firstly by Little of about 60 year ago [77].

The 1D-superconductivity in the "excitonic model" of Little ([77], see also [78] and references therein) is based on charge oscillations in the side chains, which provoke an attractive interaction between electrons moving in a long chain. We suggest, that this new plasmon-phonon scenario of electron-electron pairing proposed here for ZrB$_{12}$, may be valid also in the case of (Y,La)H$_n$ polyhydrides with labile H$_n$ clusters, which are the superconductors with the highest transition temperature at present $T_c$ ~250-270 K (see, for example, [56-58]), and for high-T$_c$ cuprates with collective dynamics of O$_6$ octahedra centered by Cu-ion.

Finally, in this section we analyze the features of the charge transport and thermodynamic characteristics using the well-known relation

$$S = \frac{C_{ph}}{ne} \bigg/ \left(1 + \tau_{e-ph}/\tau\right) \qquad (3),$$

which connects the Hall coefficient $R_H = 1/ne$ ($n$ is the concentration of charge carriers, $e$ is the electron charge) with phonon component of thermopower and phonon contribution to the heat capacity $C_{ph} = C - \gamma T$ [67] (in Eq. (3) $\tau_{e-ph}$ and $\tau$ denote the electron-phonon relaxation time and the relaxation time of the phonon gas, respectively). Using the experimental results shown in Figs. 9, 11, 14 and Hall effect measurements of LuB$_{12}$ [26, 27, 44], we are able to estimate the temperature dependence of the factor $(1+\tau_{e-p}/\tau)^{-1} \approx \tau/\tau_{e-p} \approx S/((C-\gamma T)R_H)$, which determines the relative change of relaxation times of the phonon gas and conduction electrons. It can be seen in Fig. 16a that the ratio $\tau/\tau_{e-p}$ of relaxation times in ZrB$_{12}$ demonstrates power-law behavior $\tau/\tau_{e-p}(T) \sim T^{-\alpha}$ both in the interval II ($\alpha \approx 5$) and III-IV ($\alpha \approx 1$). In the case of LuB$_{12}$ we define similar behavior $\tau/\tau_{e-p}(T) \sim T^{-\alpha}$ with $\alpha \approx 1$ above $T^* \approx 60$ K (Fig. 16a).To separate the temperature dependences $\tau(T)$ and $\tau_{e-p}(T)$ we analyze also the thermal conductivity $\kappa(T)$ and phonon heat capacity $C-\gamma T$ in the framework of relation

$$\kappa(T) = 1/3(C-\gamma T)\cdot v_F^2 \cdot \tau(T) \quad (4),$$

where $v_F(ZrB_{12}) \approx 3.5\times 10^7$ cm/s and $v_F(LuB_{12}) \approx 7.3\times 10^7$ cm/s are Fermi velocities found for $ZrB_{12}$ from the study of superconductivity [48] and Raman measurements of $LuB_{12}$ [79], correspondingly.

When discussing the exponents with $\alpha=1$ and $\alpha=5$ detected in $ZrB_{12}$ (Fig. 16b), it is worth noting that a strong power-law dependence of $(1+\tau_{e-ph}/\tau)^{-1} \sim T^{-5}$ was observed previously in the other non-equilibrium superconductor $YB_6$ [80], and it was attributed to the regime of phonon-phonon scattering $\tau \sim T^{-5}$ [81] in the low temperature cage-glass state. Depending on the type of the normal three-phonon scattering processes either $\tau \sim T^{-5}$, or $\tau \sim T^{-4}$ behavior should be expected [81-84], and both for $ZrB_{12}$ and $LuB_{12}$ we have rough approximation $\tau \sim T^{-4}$ (see Fig. 16b). At intermediate and high temperatures a weak dependence of the product $S/((C-\gamma T)R_H) \approx \tau/\tau_{e-ph}(T) \sim T^{-1}$ was established here for $ZrB_{12}$ and $LuB_{12}$ (intervals III-IV in Fig. 16a), which could be attributed to extremely strong scattering both in the charge carriers subsystem

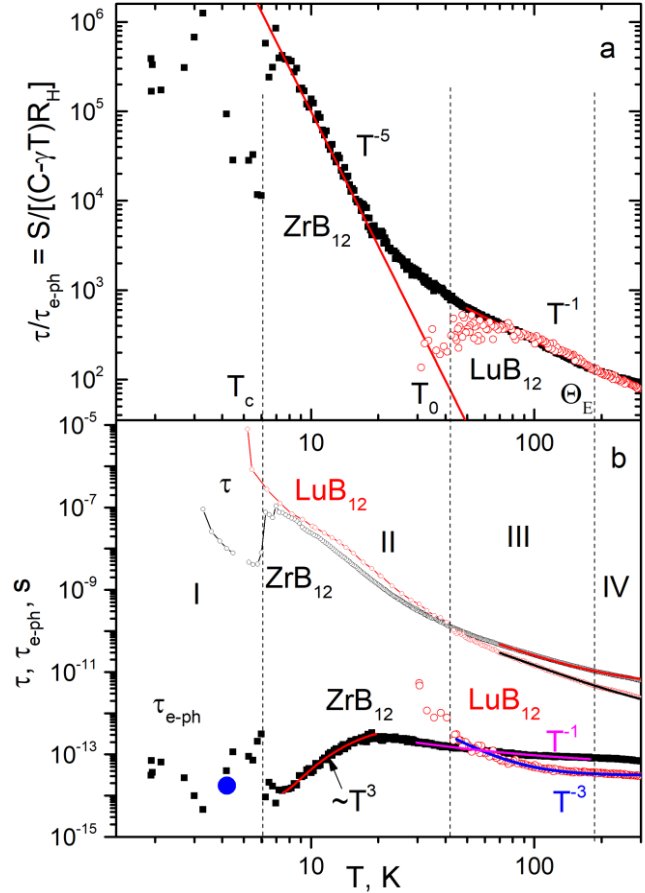

Fig. 16. Temperature dependences of the products (a) $S/[(C-\gamma T)R_H] \approx \tau/\tau_{e-ph}(T)$ (see Eq.(3)) and (b) relaxation times $\tau_{e-ph}(T)$ and $\tau(T)=3\kappa(T)/(C-\gamma T)\cdot v_F^2$ (see Eq.(4) and text) for $ZrB_{12}$ and $LuB_{12}$. Solid lines show a power-law approximation. Blue solid circle corresponds to the estimation from the analysis of quantum oscillations in $ZrB_{12}$ [73]. Roman numerals I-IV mark the temperature intervals with different phases and regimes in $ZrB_{12}$.

and in the phonon gas. It is worth noting, that the collective excitation was detected in the far infrared optical conductivity spectra of $LuB_{12}$ [76], and it was attributed to collective Jahn-Teller vibrations of $B_{12}$ clusters (ferrodistortive effect). The analysis developed in [76] allows us to conclude that about 70% of charge carriers are involved in the collective mode, producing very strong carrier scattering in the dodecaborides. Let us note also, that above $T_0$ (intervals III-IV) the ratio $\tau/\tau_{e-ph}(T)$ is almost the same for $ZrB_{12}$ and $LuB_{12}$ (Fig. 16a), but the difference appears for the $\tau(T)$ behavior, which follows to relation $\tau(T)=$

$D_0T^{-2}+\tau_0$ with quite distinct phonon gas parameters $\tau_0(ZrB_{12}) \approx 4.6 \cdot 10^{-12}$ s and $\tau_0(LuB_{12}) \approx 6.7 \cdot 10^{-13}$ s. Fig. 16b shows also the electron-phonon relaxation time $\tau_{e-ph}(T)$ detected from the same data sets $\tau/\tau_{e-p}(T)$ and $\tau(T)$. As can be seen from Fig. 16b, fast fluctuations of charge carriers in the stripes lead to very large values relaxation rates $\tau_{e-ph}^{-1}(T) \sim 10^{13}$-$10^{14}$ s both in $ZrB_{12}$ and $LuB_{12}$.

The estimates obtained here from the data analysis in framework of Eqs. (3)-(4) may be compared with the room temperature relaxation rate in lutetium dodecaboride $\tau_{JT}^{-1}(LuB_{12}) \sim 6 \cdot 10^{13}$ s$^{-1}$ [76]. Moreover, similar results have been obtained for RE and TM hexaborides ($RB_6$) with collective JT mode of rigid boron cage. The values $\tau_{JT}^{-1}(YB_6) \sim 12.6 \cdot 10^{13}$ s$^{-1}$ [85], $\tau_{JT}^{-1}(LaB_6) \sim 6.9 \cdot 10^{13}$ s$^{-1}$ [86, 87] and $\tau_{JT}^{-1}(GdB_6) \sim 10.7 \cdot 10^{13}$ s$^{-1}$ [87] are deduced in optical studies from relation $\tau_{JT}^{-1} = 2\pi\gamma_{peak}$, where $\gamma_{peak}$ is the mode damping, for non−equilibrium charge carriers participated in collective modes in $LuB_{12}$, $YB_6$, $LaB_6$ and $GdB_6$, respectively. Let us point out the dramatic difference between $\tau(T)$ and $\tau_{e-ph}(T)$ in $ZrB_{12}$ at low temperatures (see Figs. 16a-b) which allows concluding in favor of formation of strongly non-equilibrium many-body states preceding the emergence of two-gap superconductivity in this model compound with dynamic charge stripes, s-CDW and pseudo-gap. Note also that similar relaxation behavior $\tau_{e-ph} \sim T^3$ observed in $CeB_6$, where the transition to inhomogeneous magnetic phase with nanoscale spin droplets (ferrons) has been found very recently [88].

## CONCLUSIONS

We present a review of previous long-term studies of two superconductors, $LuB_{12}$ ($T_c \sim 0.4$ K, $H_c \sim 20$ Oe) and $ZrB_{12}$ ($T_c \sim 6$ K, $H_c \sim 400$ Oe), which are supplemented by new results of charge transport (magnetoresistance, Hall and Seebeck coefficients), thermal conductivity and heat capacity measurements combined with precise X-ray diffraction investigation of the crystal and electronic structure. The electron density distribution, determined by the maximum entropy method in these two dodecaborides with similar conduction bands and phonon spectra, demonstrates two main differences: (*i*) a sub-structural CDW with triangular lattice is clearly observed in $ZrB_{12}$ and not essential in $LuB_{12}$, and (*ii*) grids of dynamic charge stripes are established, located predominantly on the 2*p*-states of boron clusters in $ZrB_{12}$, instead of the unique 5*d*-2*p* Lu and B linear filamentary structure, that has been found at low temperatures in $LuB_{12}$. As a result, the nanoscale electron phase separation caused by the Jahn-Teller lattice instability of the rigid boron framework induces in $ZrB_{12}$ transverse to the stripes quasi-local vibrations of $Zr^{4+}$ ions pairs, and the Einstein phonon modes that mediate the superconductivity turn out to be synchronized by the quasi-1D collective dynamics of boron chains. We propose, that the attraction between electrons located in Zr-Zr pairs are mediated by both the quasi-local oscillations of Zr-ions (*phonons*) and by high frequency collective JT-vibrations (*plasmons*). This leads to a composite *plasmon-phonon pairing mechanism*, which includes two quasi-local vibrations of Zr-ions separated by the superconducting coherence length of 570 Å and synchronized by collective 1D-oscillations of $B_{12}$ clusters, and the scenario may be common to different classes of high-$T_c$ superconductors. The low-frequency vibrations of Lu ions in 5*d*-2*p* stripes are longitudinal, so the proposed plasmon-phonon mechanism is not working here. Additionally, two phase transitions in $ZrB_{12}$, at $T_0 \approx 42$ K and $T \sim \Theta_E \approx 180$ K, are discussed. They can be attributed to changes in the configuration and the pinning of sliding CDWs.

**Acknowledgements**

The work was partly performed using the equipment of the Shared Research Centers of FSRC 'Crystallography and Photonics' of NRC KI and Lebedev Physical Institute of RAS. The work of N. B. Bolotina and O. N. Khrykina in the part of structural analysis was performed within the framework of the State Assignment of NRC 'Kurchatov Institute'. The authors are grateful to S. V. Demishev and V. V. Glushkov for useful discussions. S.G. and K.F. acknowledge the support of the Slovak Research and Development Agency under contracts No. APVV−23−0226 and VEGA 2/0034/24.

## REFERENCES

1. B. T. Matthias, T. H. Geballe, K. Andres, E. Corenzwit, G. W. Hull, and J. P. Maita, "Superconductivity and Antiferromagnetism in Boron-Rich Latfices", Science **159**, 530 (1968).

2.  I. Bat'ko, M. Bat'kova, K. Flachbart, V.B. Filippov, Yu.B. Paderno, N.Yu. Shicevalova, Th. Wagner, "Electrical resistivity and superconductivity of LaB$_6$ and LuB$_{12}$", Journal of Alloys and Compounds **217**, L1-L3 (1995).

3.  K. Flachbart, S. Gabani, K. Gloos, M. Meissner, M. Opel, Y. Paderno, V. Pavlık, P. Samuely, E. Schuberth, N. Shitsevalova, K. Siemensmeyer, and P. Szabo, "Low Temperature Properties and Superconductivity of LuB$_{12}$", Journal of Low Temperature Physics, **140** (5/6), 339-353 (2005).

4.  A. Leithe-Jasper, A. Sato and T. Tanaka, "Refinement of the crystal structure of zirconium dodecaboride, ZrB$_{12}$, at 140 K and 293 K", Z. Kristallogr. New Crystal Structures, **217**, 319-320, (2002).

5.  W.N. Lipscomb, and D. Britton, "Valence Structure of the Higher Borides", J. Chem. Phys., **33**, 275 (1960).

6.  R.W. Johnson and A.H. Daane, "Electron Requirements of Bonds in Metal Borides", J. Chem. Phys., **38**, 425 (1963).

7.  I.R. Shein, A.L. Ivanovskii, "Band structure of superconducting dodecaborides YB$_{12}$ and ZrB$_{12}$", Phys. Solid State **45**, 1429–1434 (2003).

8.  B Jäger, S Paluch, O J Żogał, W Wolf, P Herzig, V B Filippov, N Shitsevalova and Y Paderno," Characterization of the electronic properties of YB$_{12}$, ZrB$_{12}$, and LuB$_{12}$ using $^{11}$B NMR and first-principles calculations", J. Phys.: Condens. Matter **18**, 2525–2535, (2006).

9.  G. E. Grechnev, A. E. Baranovskiy, V. D. Fil, T. V. Ignatova, I. G. Kolobov, A. V. Logosha, N. Yu. Shitsevalova, V. B. Filippov and Olle Eriksson, "Electronic structure and bulk properties of MB$_6$ and MB$_{12}$ borides", Low Temperature Physics **34**, 921-929 (2008).

10. R. Lortz, Y. Wang, S. Abe, C. Meingast, Y. B. Paderno, V. Filippov, and A. Junod, "Specific heat, magnetic susceptibility, resistivity and thermal expansion of the superconductor ZrB$_{12}$", Phys. Rev. B **72**, 024547 (2005).

11. A. V. Rybina, K. S. Nemkovski, P. A. Alekseev, J. M. Mignot, E. S. Clementyev, M. Johnson, L. Capogna, A. V. Dukhnenko, A. B. Lyashenko, and V. B. Filippov, "Lattice dynamics in ZrB$_{12}$ and LuB$_{12}$: Ab initio calculations and inelastic neutron scattering measurements" // Phys. Rev. B **82**, 024302, (2010).

12. A. V. Rybina, K. S. Nemkovski, V. B. Filipov, and A. V. Dukhnenko, "Phonons in ZrB$_{12}$", Phys. Solid State **52** (5), 894 - 898 (2010).

13. D. Mandrus, B. C. Sales, and R. Jin, "Localized vibrational mode analysis of the resistivity and specific heat of LaB$_6$", Phys. Rev. B **64**, 012302 (2001).

14. V. A. Gasparov, N. S. Sidorov, and I. I. Zver'kova, "Two-gap superconductivity in ZrB$_{12}$: Temperature dependence of critical magnetic fields in single crystals", Phys. Rev. B **73**, 094510 (2006).

15. V. Glushkov, M. Ignatov, S. Demishev, V. Filippov, K. Flachbart, T. Ishchenko, A. Kuznetsov, N. Samarin, N. Shitsevalova, and N. Sluchanko, "Phonon drag induced by Einstein mode in ZrB$_{12}$", Phys. Stat. Sol. b **243** (11), R72–R74 (2006).

16. A. Bouvet, T. Kasuya, M. Bonnet, L. P. Regnault, J. Rossat_Mignod, F. Iga, B. Fek, and A. Severing, "Magnetic excitations observed by means of inelastic neutron scattering in polycrystalline YbB$_{12}$", J. Phys.: Condens. Matter **10**, 5667 – 5677, (1998).

17. A. V. Rybina, P. A. Alekseev, J. M. Mignot, E. V. Nefeodova, K. S. Nemkovski, R. I. Bewley, N. Yu. Shitsevalova, Yu. B. Paderno, F. Iga, and T. Takabatake, "Lattice dynamics and magneto-elastic coupling in Kondo-insulator YbB$_{12}$", J. Phys.: Conf. Ser. **92**, 012074 (2007).

18. A. Czopnik, N. Shitsevalova, A. Krivchikov, V. Pluzhnikov, Y. Paderno, and Y. Onuki, "Thermal properties of rare earth dodecaborides", J. Solid State Chem. **177**, 507 - 514 (2004).

19. A. Czopnik, N. Shitsevalova, V. Pluzhnikov, A. Krivchikov, Yu. Paderno, and Y. Onuki, "Low-temperature thermal properties of yttrium and lutetium dodecaborides", J. Phys.: Condens. Matter **17**, 5971 – 5985, (2005).

20. J. Teyssier, R. Lortz, A. Petrovic, D. van der Marel, V. Filippov and N. Shitsevalova, "Effect of electron-phonon coupling on the superconducting transition temperature in dodecaboride superconductors: A comparison of $LuB_{12}$ with $ZrB_{12}$", Physical Review B **78**, 134504 (2008).

21. H. Werheit, Yu. Paderno, V. Filippov, V. Paderno, A. Pietraszko, M. Armbrüster, U. Schwarz, "Peculiarities in the Raman spectra of $ZrB_{12}$ and $LuB_{12}$ single crystals", Journal of Solid State Chemistry **179**, 2761–2767, (2006).

22. J. Teyssier, A. B. Kuzmenko, D. van der Marel, F. Marsiglio, A. B. Liashchenko, N. Shitsevalova, and V. Filippov, "Optical study of electronic structure and electron-phonon coupling in $ZrB_{12}$", Physical Review B **75**, 134503, (2007).

23. N. E. Sluchanko, A. N. Azarevich, A. V. Bogach, I. I. Vlasov, V. V. Glushkov, S. V. Demishev, A. A. Maksimov, I. I. Tartakovskii, E. V. Filatov, K. Flachbart, S. Gabani, V. B. Filippov, N. Yu. Shitsevalova, and V. V. Moshchalkov, "Effects of Disorder and Isotopic Substitution in the Specific Heat and Raman Scattering in $LuB_{12}$", Journal of Experimental and Theoretical Physics **113** (3), 468–482, (2011).

24. N. Sluchanko, S. Gavrilkin, K. Mitsen, A. Kuznetsov, I. Sannikov, V. Glushkov, S. Demishev, A. Azarevich, A. Bogach, A. Lyashenko, A. Dukhnenko, V. Filipov, S. Gabani, K. Flachbart, J. Vanacken, Gufei Zhang, V. Moshchalkov, "Superconductivity in $ZrB_{12}$ and $LuB_{12}$ with Various Boron Isotopes", J Supercond Nov Magn **26**, 1663–1667, (2013).

25. N. E. Sluchanko, A. N. Azarevich, M. A. Anisimov, A. V. Bogach, S. Yu. Gavrilkin, V. V. Glushkov, S. V. Demishev, A. A. Maksimov, I. I. Tartakovskii, E. V. Filatov, V. B. Filippov, and A. B. Lyashchenko, "Raman Scattering in $ZrB_{12}$ Cage Glass", JETP Letters **103** (11), 674–679, (2016).

26. N. E. Sluchanko, A. N. Azarevich, M. A. Anisimov, A. V. Bogach, S. Yu. Gavrilkin, M. I. Gilmanov, V. V. Glushkov, S. V. Demishev, A. L. Khoroshilov, A. V. Dukhnenko, K. V. Mitsen, N. Yu. Shitsevalova, V. B. Filippov, V. V. Voronov, and K. Flachbart, "Suppression of superconductivity in $Lu_xZr_{1-x}B_{12}$: Evidence of static magnetic moments induced by nonmagnetic impurities", Phys. Rev. B **93**, 085130, (2016).

27. N. E. Sluchanko, A. N. Azarevich, A. V. Bogach, V. V. Glushkov, S. V. Demishev, A. V. Kuznetsov, K. S. Lyubshov, V. B. Filippov, and N. Yu. Shitsevalova, "Isotope Effect in Charge Transport of $LuB_{12}$", JETP **111** (2), 279 – 284, (2010).

28. N. Sluchanko, L. Bogomolov, V. Glushkov, S. Demishev, M. Ignatov, Eu. Khayrullin, N. Samarin, D. Sluchanko, A. Levchenko, N. Shitsevalova, and K. Flachbart, "Anomalous charge transport in $RB_{12}$ (R = Ho, Er, Tm, Lu)", Phys. Status Solidi B **243**, R63 – R65, (2006).

29. N. E. Sluchanko, A. V. Bogach, V. V. Glushkov, S. V. Demishev, N. A. Samarin, D. N. Sluchanko, A. V. Dukhnenko, and A. V. Levchenko, "Anomalies of Magnetoresistance of Compounds with Atomic Clusters $RB_{12}$ (R = Ho, Er, Tm, Lu)", JETP **108** (4), 668 - 687, (2009).

30. R. Franz, and H. Werheit, "Jahn-Teller effect of the $B_{12}$ icosahedron and its general influence on the valence band structures of boron-rich solids", Europhys. Lett. **9** (2), 145–150, (1989).

31. R. Franz, and H. Werheit, "Influence of the Jahn-Teller effect on the electronic band structure of boron-rich solids containing $B_{12}$ icosahedra", AIP Conf. Proc. **231**, 29–36 (1991).

32. N. Sluchanko, A. Bogach, N. Bolotina, V. Glushkov, S. Demishev, A. Dudka, V. Krasnorussky, O. Khrykina, K. Krasikov, V. Mironov, V. B. Filipov, and N. Shitsevalova, "Rattling mode and symmetry lowering resulting from the instability of the $B_{12}$ molecule in $LuB_{12}$", Phys. Rev. B **97**, 035150 (2018).

33. G. A. Gehring and K. A. Gehring, "Co-operative Jahn-Teller effects", Rep. Prog. Phys. **38**, 1–89, (1975).

34. I. B. Bersuker and V. Z. Polinger (eds.), "Vibronic interactions in molecules and crystals", Springer Series in Chemical Physics **49**, (Springer, Berlin, Heidelberg, 1989)

35. M. D. Kaplan and B. G. Vekhter, *Cooperative phenomena in Jahn-Teller crystals* (Plenum Press, New York, 1995).

36. I. Bersuker, *The Jahn-Teller Effect* (Cambridge University Press, Cambridge, 2006).

37. K. Flachbart, P. Alekseev, G. Grechnev, N. Shitsevalova, K. Siemensmeyer, N. Sluchanko and O. Zogal, *Rare Earths: Research and Applications* ed. K. N. Delfrey (Hauppauge, NY: Nova Science), pp 79–125 (2008).

38. A. P. Dudka, I. A. Verin, E. S. Smirnova, "Calibration of Cryojet and Cobra Plus cryosystems used in X-ray diffraction studies", Crystallogr. Rep. **61**, 692–696 (2016).

39. A.P. Dudka, O.N. Khrykina, N.B. Bolotina, N.Yu. Shitsevalova, "Jahn–Teller lattice distortions and asymmetric electron density distribution in the structure of $TmB_{12}$ dodecaboride in the temperature range of 85–293 K", Crystallography Reports **64** (5), 737–742, (2019).

40. N.B. Bolotina, A.P. Dudka, O.N. Khrykina, V.N. Krasnorussky, N.Yu. Shitsevalova, V.B. Filipov, N.E. Sluchanko, "The lower symmetry electron-density distribution and the charge transport anisotropy in cubic dodecaboride $LuB_{12}$", J. Phys.: Condens. Matter **30**, 265402, (2018).

41. N.B. Bolotina, A.P. Dudka, O.N. Khrykina, V.V. Glushkov, A.N. Azarevich, V.N. Krasnorussky, S. Gabani, N.Yu. Shitsevalova, A.V. Dukhnenko, V.B. Filipov, N.E. Sluchanko, "On the role of isotopic composition in crystal structure, thermal and charge-transport characteristics of dodecaborides $Lu^N B_{12}$ with the Jahn-Teller instability", J. Physics and Chemistry of Solids **129**, 434–441, (2019).

42. O. N. Khrykina, A. P. Dudka, N. B. Bolotina, N. E. Sluchanko, N. Yu. Shitsevalova, "Structural instability and poorly defined phase transitions in rare-earth dodecaborides $RB_{12}$ (R=Ho-Lu) at intermediate temperatures", Solid State Sciences **107**, 106273, (2020).

43. N. B. Bolotina, A. P. Dudka, O. N. Khrykina, V.S. Mironov, "Crystal structure of dodecaborides: complexity in simplicity", in *Rare-Earth Borides*, edited by D. S. Inosov (Jenny Stanford Publishing, Singapore, 2021), Chap.3, pp. 293-330.

44. N. Sluchanko, A. Azarevich, A. Bogach, S. Demishev, K. Krasikov, V. Voronov, V. Filipov, N. Shitsevalova, V. Glushkov, "Hall effect and symmetry breaking in non-magnetic metal with dynamic charge stripes", Phys. Rev. B **103**, 035117, (2021).

45. T. S. Altshuler, Y. V. Goryunov, M. S. Bresler, F. Iga, T. Takabatake, "Ion pairs and spontaneous break of symmetry in the valence-fluctuating compound $YbB_{12}$" // Phys. Rev. B **68**, 014425, (2003).

46. N. Bolotina, O. Khrykina, A. Azarevich, S. Gavrilkin, and N. Sluchanko, "Fine details of crystal structure and atomic vibrations in $YbB_{12}$ with a metal–insulator transition", Acta Cryst. B **76**, 1117–1127, (2020).

47. B. Z. Malkin, E. A. Goremychkin, K. Siemensmeyer, S. Gabáni, K. Flachbart, M. Rajňák, A. L. Khoroshilov, K. M. Krasikov, N. Yu. Shitsevalova, V. B. Filipov, and N. E. Sluchanko, "Crystal-field potential and short-range order effects in inelastic neutron scattering, magnetization, and heat capacity of the cage-glass compound HoB12", Phys. Rev. B **104**, 134436, (2021).

48. A. Azarevich, A. Bogach, V. Glushkov, S. Demishev, A. Khoroshilov, K. Krasikov, V. Voronov, N. Shitsevalova, V. Filipov, S. Gabáni, K. Flachbart, A. Kuznetsov, S. Gavrilkin, K. Mitsen, S. J. Blundell, N. E. Sluchanko, "Inhomogeneous superconductivity in $Lu_xZr_{1-x}B_{12}$ dodecaborides with dynamic charge stripes", Phys. Rev. B **103**, 104515, (2021).

49. F. K. K. Kirschner, N. E. Sluchanko, V. B. Filipov, F. L. Pratt, C. Baines, N. Yu. Shitsevalova, and S. J. Blundell, "Observation of a crossover from nodal to gapped superconductivity in $Lu_xZr_{1-x}B_{12}$", Phys. Rev. B **98**, 094505, (2018).

50. N. B. Bolotina, O. N. Khrykina, A. N. Azarevich, N. Yu. Shitsevalova, V. B. Filipov, S. Yu. Gavrilkin, K. V. Mitsen, N. E. Sluchanko, "Checkerboard patterns of charge stripes in a two-gap superconductor $ZrB_{12}$", Phys. Rev B **105**, 054511, (2022).

51. P. K. Biswas, F. N. Rybakov, R. P. Singh, Saumya Mukherjee, N. Parzyk, G. Balakrishnan, M. R. Lees, C. D. Dewhurst, E. Babaev, A. D. Hillier, and D. Mc K. Paul, "Coexistence of type-I and type-II superconductivity signatures in $ZrB_{12}$ probed by muon spin rotation measurements", Phys. Rev. B **102**, 144523, (2020).

52. S. Datta, S. Howlader, A. R. Prakash Singh, G. Sheet, "Anisotropic superconductivity in $ZrB_{12}$ near the critical Bogomolnyi point", Phys. Rev. B **105**, 094504, (2022)

53. N.B. Bolotina, O.N. Khrykina, A.N. Azarevich, N.Yu. Shitsevalova, V.B. Filipov, S.Yu. Gavrilkin, A.Yu. Tsvetkov, S. Gabáni, K. Flachbart, V.V. Voronov, N.E. Sluchanko, "Low temperature singularities of electron density in a two-gap superconductor $ZrB_{12}$", Solid St. Sci. **142**, 107245, (2023).

54. S. Thakur et al., D. Biswas, N. Sahadev, P. K. Biswas, G. Balakrishnan, K. Maiti, "Complex spectral evolution in a BCS superconductor $ZrB_{12}$", Sci. Rep. **3**, 3342, (2013).

55. A. Chubukov, D. Pines, J. Schmalian, "A Spin Fluctuation for dwave Superconductivity. In The Physics of Superconductors", Vol. I: *Conventional and High-Tc Superconductors*, K-H. Benneman, J. B. Ketterson, Eds. (Springer: Berlin, 2003), pp 1349- 1407.

56. I. A. Troyan, D. V. Semenok, A. G. Kvashnin, A. V. Sadakov, O. A. Sobolevskiy, V. M. Pudalov, A. G. Ivanova, V. B. Prakapenka, E. Greenberg, A. G. Gavriliuk, I. S. Lyubutin, V. V. Struzhkin, A. Bergara, I. Errea, R. Bianco, M. Calandra, F. Mauri, L. Monacelli, R. Akashi, and A. R. Oganov, "Anomalous High-Temperature Superconductivity in $YH_6$", Adv. Mater. **33**, 2006832, (2021).

57. I. A. Troyan, D. V. Semenok, A. G. Ivanova, A. G. Kvashnin, D. Zhou, A. V. Sadakov, O. A. Sobolevskiy, V. M. Pudalov, I. S. Lyubutin, A. R. Oganov, "High-temperature superconductivity in hydrides", UFN **192**, 799–813, (2022).

58. D. Sun, V. S. Minkov, S. Mozaffari, Y. Sun, Y. Ma, S. Chariton, V. B. Prakapenka, M. I. Eremets, L. Balicas, F. F. Balakirev, "High-temperature superconductivity on the verge of a structural instability in lanthanum superhydride", Nat. Comm. **12**, 6863, (2021).

59. N. Shitsevalova, "Crystal chemistry and crystal growth of rare-earth borides", in: edited by D. S. Inosov, *Rare-Earth Borides*, (Jenny Stanford Publishing, Singapore, 2021), Chap. 1, pp. 1– 243

60. N. E. Sluchanko, A. N. Azarevich, A. V. Bogach, V. V. Glushkov, S. V. Demishev, M. A. Anisimov, A. V. Levchenko, V. B. Filipov, N. Yu. Shitsevalova, "Hall and Transverse Even Effects in the Vicinity of a Quantum Critical Point in $Tm_{1-x}Yb_xB_{12}$", J. Exp. Theor. Phys. **115**, 509–526, (2012).

61. N. E. Sluchanko, A. N. Azarevich, A. V. Bogach, N. B. Bolotina, V. V. Glushkov, S. V. Demishev, A. P. Dudka, O. N. Khrykina, V. B. Filipov, N. Yu. Shitsevalova, G. A. Komandin, A. V. Muratov, Yu. A. Aleshchenko, E. S. Zhukova, B. P. Gorshunov, "Observation of dynamic charge stripes at the metal-insulator transition in $Tm_{0.19}Yb_{0.81}B_{12}$", J. Phys.: Condens. Matter **31**, 065604, (2019).

62. V. Petříček, M. Dušek, L. Palatinus, "Crystallographic Computing System JANA2006: General Features", Z. Kristallogr. **229**, 345–352, (2014).

63. A. P. Dudka, N. B. Bolotina, O. N. Khrykina, "DebyeFit: a simple tool to get an appropriate model of atomic vibrations in solids from atomic displacement parameters obtained at different temperatures", J. Appl. Cryst. **52**, 690–692, (2019).

64. Y. Paderno, A. Liashchenko, V. Filippov, and A. Dukhnenko, in *Science for Materials*, "the Frontier of Centuries Advantages and Challenges", Conf. Proc. IPMS NASU, edited by V. Skorokhod, Kiev, 2002, p. 404.

65. Ma, T., L, H., Zheng, X., Wang, Sh., Wang, X., Zhao, H., Han, S., Liu, J., Zhang, R., Zhu, P., Long, Y., Cheng J., Ma, Y., Zhao, Yu., Jin, Ch. & Yu, Xiaohui, "Ultrastrong Boron Frameworks in $ZrB_{12}$: A Highway for Electron Conducting", Adv. Mater. **29**, 1604003, (2017).

66. J.R. Cooper, "Electrical resistivity of an Einstein solid", Phys. Rev. B **9**, 2778, (1974).

67. P.M. Chaikin, In: *Organic superconductivity*, ed. by V.Z. Kresin and W.A. Little, Plenum Press, New York (1990), p.101.

68. Yu.I. Latyshev, A.P. Orlov, A.Yu. Latyshev, A.-M. Smolovich, P. Monceau, and D. Vignolles, "Recent experiments on interlayer tunneling spectroscopy and transverse electric field effect in $NbSe_3$", Physica B **404**, 399-403, (2009).

69. H Misiorek, J Mucha, A Jezowski, Y Paderno and N Shitsevalova, "Thermal conductivity of rare-earth element dodecaborides", J. Phys.: Condens. Matter **7**, 8927, (1995).

70. O.M. Vyaselev, A.A. Gippius, N.E. Sluchanko, N.Yu. Shitsevalova, "Low-Temperature Crystallization of Structural Defects in $LuB_{12}$ According to $^{175}Lu$ NMR Data" // JETP Letters **119**, 529–533, (2024).

71. P Chandra, "Fluctuation effects on the Pauli susceptibility at a Peierls transition", J. Phys.: Condens. Matter **1**, 10067 -10080, (1989).

72. K. M. Krasikov, A. N. Azarevich, V. V. Glushkov, S. V. Demishev, A. L. Khoroshilov, A. V. Bogach, N. Yu. Shitsevalova, V. B. Filippov, and N. E. Sluchanko, "Breaking of Cubic Symmetry in Rare-Earth Dodecaborides with Dynamic Charge Stripes", JETP Letters **112**, 415–421, (2020).

73. V. A. Gasparov, I. Sheikin, F. Levy, J. Teyssier, G. Santi, "Study of the Fermi Surface of $ZrB_{12}$ Using the de Haas-van Alphen Effect", Phys. Rev. Lett. **101**, 097006, (2008).

74. Chih-Wei Chen, Jesse Choe and E Morosan, "Charge density waves in strongly correlated electron systems", Rep. Prog. Phys. **79**, 084505, (2016).

75. Xuetao Zhu, Jiandong Guo, Jiandi Zhang, E. W. Plummer, "Misconceptions associated with the origin of charge density waves", Advances in Physics: X **2** (3), 622-640, (2017).

76. B. P. Gorshunov, E. S. Zhukova, G. A. Komandin, V. I. Torgashev, A. V. Muratov, Yu. A. Aleshchenko, S. V. Demishev, N. Yu. Shitsevalova, V. B. Filipov N. E. Sluchanko, "Collective Infrared Excitation in $LuB_{12}$ Cage-Glass", JETP Letters **107**, 100, (2018).

77. W.A. Little, "Possibility of Synthesizing an Organic Superconductor", Phys. Rev. **134**, A1416 (1964).

78. *High-Temperature Superconductivity*, Eds. V.L. Ginzburg, D.A. Kirzhnits (New York: Consultant Bureau, 1982)

79. Yu. S. Ponosov, A.A. Makhnev, S.V. Streltsov, V.B. Filipov, N. Yu. Shitsevalova, "Raman study of coupled electronic and phononic excitations in $LuB_{12}$", Journal of Alloys and Compounds **704**, 390-397, (2017).

80. N. Sluchanko, V. Glushkov, S. Demishev, A. Azarevich, M. Anisimov, A. Bogach, V. Voronov, S. Gavrilkin, K. Mitsen, A. Kuznetsov, I. Sannikov, N. Shitsevalova, Filipov, M. Kondrin, S. Gabáni and K. Flachbart, "Lattice instability and enhancement of superconductivity in $YB_6$", Phys. Rev. B **96**, 144501, (2017).

81. A.M. Kosevich, *Theory of Crystal Lattice*, Kharkov State University, Vishcha Shkola, Kharkov, 1988 (in Russian); Translated into English: WILEY-VCH, Berlin, New York, 1999

82. A. Anselm, *Introduction to Semiconductor Theory*, Translated from the Russian. (Mir, Moscow, 1981). 645 p

83. S. Simons, "On the Mutual Interaction of Parallel Phonons", Proc. Phys. Soc. **82**, 401, (1963).

84. C. Herring, "Role of Low-Energy Phonons in Thermal Conduction", Phys. Rev. **95**, 954, (1954).

85. N.E. Sluchanko, E.S. Zhukova, L.N. Alyabyeva, B.P. Gorshunov, A.V. Muratov, Yu.A. Aleshchenko, A.N. Azarevich, M.A. Anisimov, N.Yu. Shitsevalova, S.E. Polovets, and V.B. Filipov, "Collective and quasi-local modes in the optical spectra of $YB_6$ and $YbB_6$ hexaborides with Jahn–Teller structural instability", J. Exp. Theor. Phys. **136**, 148 (2023).

86. E.S. Zhukova, B.P. Gorshunov, M. Dressel, G.A. Komandin, M.A. Belyanchikov, Z.V. Bedran, A.V. Muratov, Y.A. Aleshchenko, M.A. Anisimov, N.Yu. Shitsevalova, A.V. Dukhnenko, V.B. Filipov, V.V. Voronov, and N.E. Sluchanko, "Boron $^{10}B$-$^{11}B$ isotope substitution as a probe of the mechanism responsible for the record thermionic emission in $LaB_6$ with the Jahn-Teller instability", JETP Lett. **110**, 79, (2019).

87. E.S. Zhukova, B.P. Gorshunov, G.A. Komandin, L.N. Alyabyeva, A.V. Muratov, Yu.A. Aleshchenko, M.A. Anisimov, N.Yu. Shitsevalova, S.E. Polovets, V.B. Filipov, V.V. Voronov, and N.E. Sluchanko, "Collective infrared excitation in rare−earth $Gd_xLa_{1-x}B_6$ hexaborides", Phys. Rev. B **100**, 104302, (2019).

88. A.N. Azarevich, O.N. Khrykina, N.B. Bolotina, V.G. Gridchina, A.V. Bogach, S.V. Demishev, V.N. Krasnorussky, S. Yu. Gavrilkin, A.Yu. Tsvetkov, N.Yu. Shitsevalova, V.V. Voronov, K.I. Kugel, A.L. Rakhmanov, S. Gabáni, K. Flachbart, N.E. Sluchanko, "Evidence for spin droplets (ferrons) formation in the heavy fermion metal $CeB_6$ with dynamic charge stripes", Solid State Sciences, in print (2025).

# TWO-GAP SUPERCONDUCTOR ZrB$_{12}$ WITH DYNAMIC STRIPES AND CHARGE DENSITY WAVES: CRYSTAL STRUCTURE, PHYSICAL PROPERTIES AND PAIRING MECHANISM

A. N. Azarevich, N. B. Bolotina, O. N. Khrykina, A. V. Bogach, K. M. Krasikov, A. Yu. Tsvetkov, S. Yu. Gavrilkin, V. V. Voronov, S. Gabani, K. Flachbart, A. V. Kuznetsov, N. E. Sluchanko

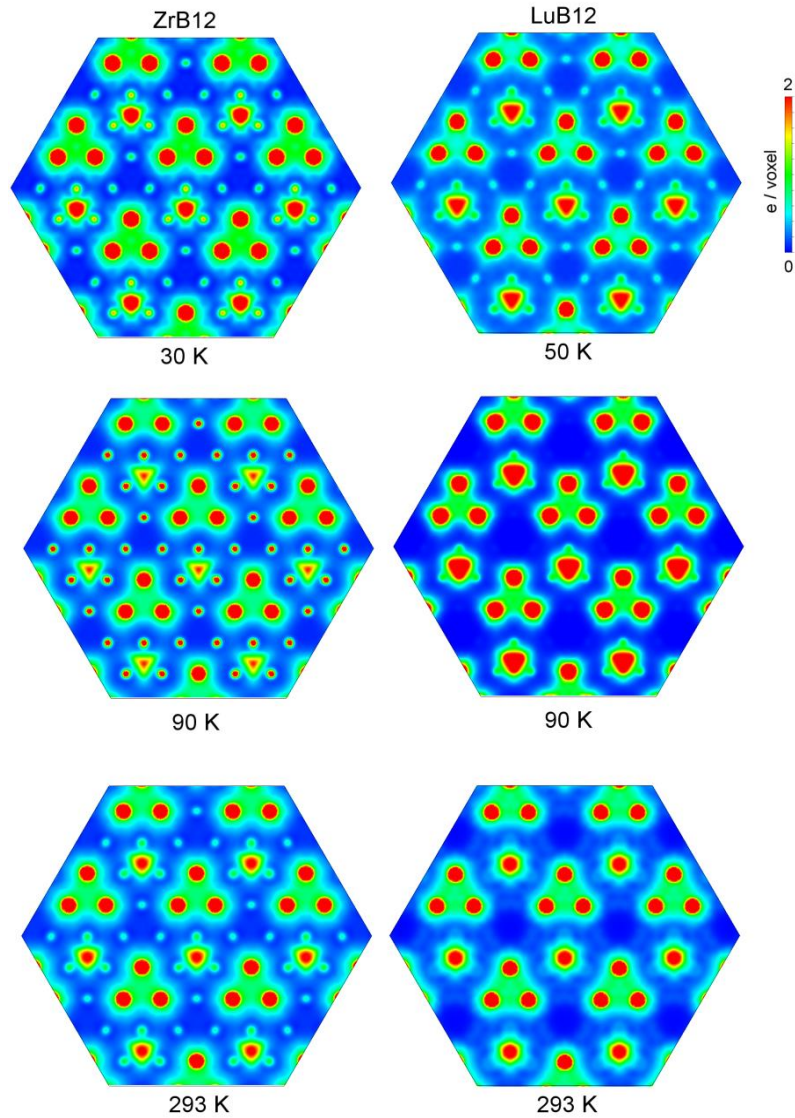

**Fig. S1**. MEM maps of ED distribution at room and low temperatures in the planes (111) of ZrB$_{12}$ and LuB$_{12}$. The calculations were carried out taking into account the cubic symmetry of the structural model. Large red circles are boron atoms in the triangular faces of B$_{12}$ cuboctahedra. Zr atoms are out of plane by ~0.7 Å. The ED peaks are cut off at a height of 2 e/voxel to show in color fine details of the ED distribution. Antinodes of the sub-structural charge density wave (s-CDW) are observed in the interstices of the crystal lattice with a period approximately equal to the distances between boron atoms, r(B – B)~1.7 Å.

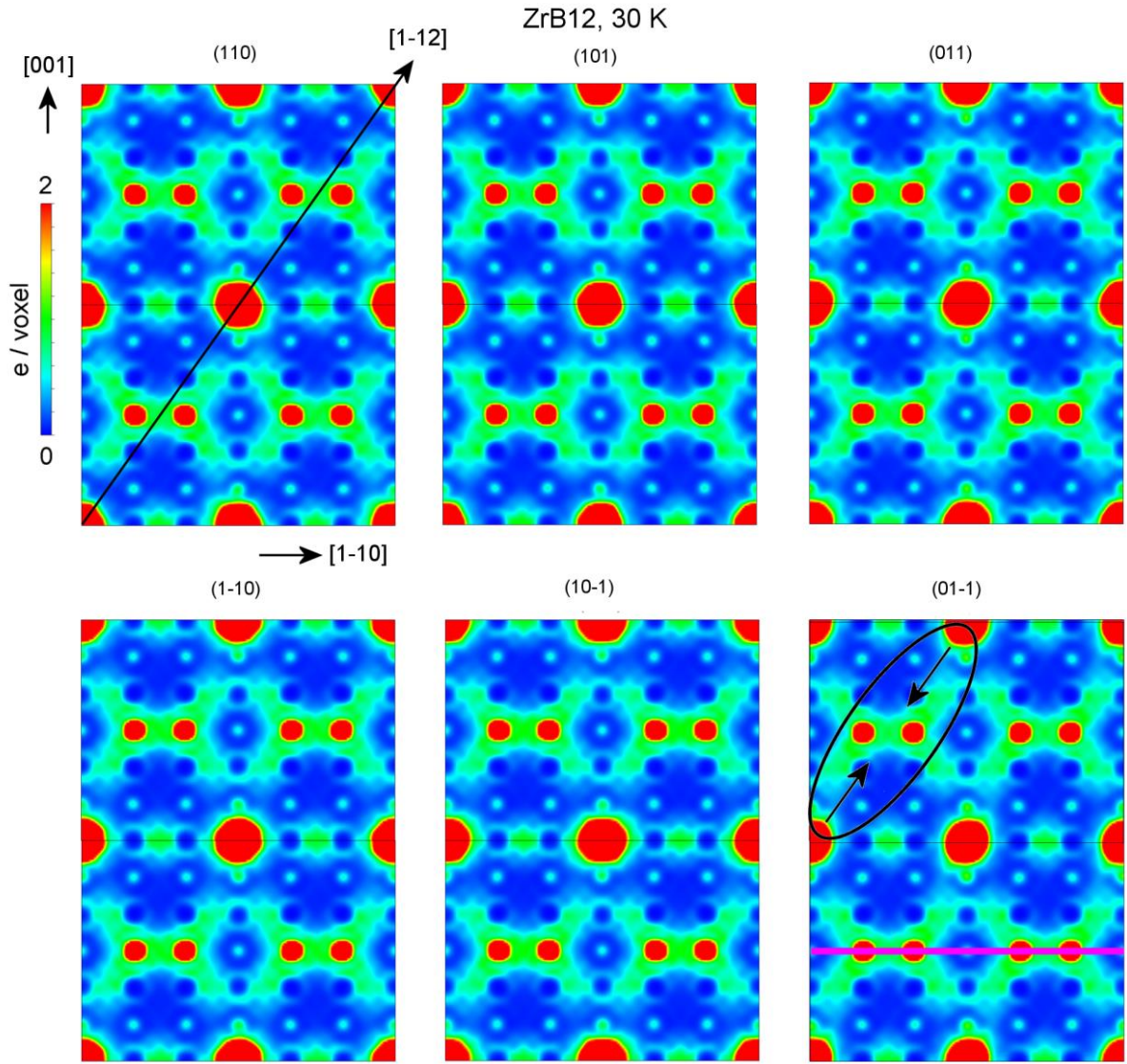

**Fig. S2**. MEM maps of the ED distribution in the family of {110} planes of ZrB$_{12}$ at $T$= 30 K. The calculations were carried out without taking into account the cubic symmetry of the structural model. The ED peaks are truncated at a height of 2 e/voxel to show in color the fine details of the ED distribution in the interstices of the crystal lattice. Each plane contains Zr atoms (large red circles) and B atoms (small red circles). In the (01-1) plane, a Zr-Zr vibrational pair formed along one of the axes of the <112> family and a stripe along one of the <110> directions in the boron lattice are shown.